# Inversion of the Impedance Response towards Physical Parameter Extraction using Interpretable Machine Learning

*Mahmoud Nabil[1], Isel Grau-García[2], Ricardo Grau-Crespo[3], Said Hamad[1] and Juan A. Anta,[1,*]*

*[1] Center for Nanoscience and Sustainable Technologies (CNATS). Department of Physical, Chemical, and Natural Systems, Universidad Pablo de Olavide, Sevilla 41013, Spain*

*[2] Information Systems Group and Eindhoven Artificial Intelligence Systems Institute, Eindhoven University of Technology, Eindhoven 5612AZ, The Netherlands.*

*[3] School of Engineering and Materials Science, Queen Mary University of London, London E1 4NS, UK*

## ABSTRACT

Interpreting the impedance response of perovskite solar cells (PSCs) is challenging due to the complex coupling of ionic and electronic motion. While drift-diffusion (DD) modelling is a reliable method, its mathematical complexity makes directly extracting physical parameters from experimental data infeasible. This work uses DD modelling to generate a large synthetic dataset of impedance spectra for a standard $TiO_2$/MAPI/spiro configuration. This dataset trains machine learning (ML) models to predict recombination and ionic parameters from impedance measurements. A Gradient Boosting Regressor, using features from a generalized equivalent circuit, showed the best performance. Interpretative analysis indicates that open-circuit impedance experiments best probe recombination losses, while short-circuit conditions are more adequate for extracting ionic features like concentrations and mobilities. The trained ML models were tested on experimental spectra, confirming the inferred physical parameters could reproduce the data. For the studied configuration, predicted ion concentrations were (1.3-3.3) × $10^{17}$ $cm^{-3}$, ion mobilities were (5-7) × $10^{-11}$ $cm^2V^{-1}s^{-1}$, and surface recombination velocities were 7-9 and 23-40 m/s. This approach provides insights into the physical information extractable from impedance measurements and paves the way for ML models to unambiguously derive efficiency-determining parameters for solar cells.

# INTRODUCTION

Solar cells containing hybrid metal halide perovskites have revolutionized photovoltaics in the last decade due to their exceptional optical and optoelectronic properties which have pushed efficiencies very close to market requirements.[1] Despite this promising advancement, many fundamental questions remain about the physico-chemical properties of these perovskites and the research field is highly active. These materials have a dual nature,[2] behaving as classical solid semiconductors - analogous to silicon - and as ionic "slow electrolytes". The interplay between ionic and electronic motion complicates their behaviour under operational conditions (for instance, by exhibiting hysteresis in their current-voltage characteristics). It also makes the interpretation of some advanced characterization techniques notoriously difficult.

One of these techniques is Electrochemical Impedance Spectroscopy (EIS), a small-perturbation method widely used to investigate emerging photovoltaic devices.[3] EIS is based on the application of a frequency-modulated, small-amplitude electrical signal (commonly in voltage) and the measurement of the corresponding frequency-modulated electrical response (commonly in current). Impedance is the perturbation-to-signal transfer function, mathematically defined as a complex number. The actual EIS response of an optoelectronic device results from the underlying resistive, capacitive and inductive processes occurring within the device. In the case of perovskite solar cells (PSC), the interpretation and analysis of EIS measurements are singularly challenging, which has led to an extensive body of literature.[4–9] The traditional way of analysing EIS data involves equivalent circuits.[5,6,9] However, the exotic behaviour of PSCs, which is further complicated by the instability and degradation of the device, especially under illumination, has prompted researchers to employ increasingly sophisticated equivalent circuit models, involving a combination of resistors, capacitors and inductors, whose true physical origin remains far from fully understood. Drift-diffusion (DD) modelling provides an alternative approach to EIS analysis and interpretation. DD equations involve processes and parameters with a clear and unambiguous meaning, such as diffusion coefficients of electronic carriers and mobile ions, recombination lifetimes, and trap, electronic, and ionic densities, among others. DD modelling has proven useful, for instance, in determining electron and hole diffusion lengths in PSCs[10,11], understanding the origin of hysteresis in the current-voltage curve[12,13], and analysing recombination losses under operational conditions.[14–16] Despite its physical "transparency", DD analysis and fitting of experimental data require defining numerous parameters, and numerically solving complex differential equations. This makes it challenging to identify the optimal parameter set for fitting or interpreting a specific experiment or device. Nevertheless, DD modelling has recently been successfully applied to explain the shape and behaviour of impedance spectra in PSC,[5,16–18] investigate ionic properties in the active layer[19,20] and establish the most representative equivalent circuit model.[21]

In this work, we make use of the powerful capabilities of the DD formalism to simulate PSC impedance spectra for given physical parameters, generating extensive synthetic datasets suitable for training machine learning (ML) models. Our approach serves two main purposes. First, by training sufficiently predictive ML models, we aim to determine which physical parameters can be reliably obtained or extracted from impedance measurements and which remain hidden in the experimental response. In other words, we seek to establish the learnable limits of impedance spectroscopy. For instance, it is crucial to determine whether EIS experiments should preferably be conducted under open circuit (OC) or short-circuit (SC) conditions for optimal parameter extraction.[20]

Second, we investigate whether ML models can effectively solve the impedance "inversion problem", that is, identifying the correct combination of physical parameters that best fits a measured spectrum, at least for those parameters that can be reliably predicted using ML models.

Machine learning techniques have already been successfully applied to advance PSC research, supporting both material discovery and the inversion problem, where trained models are used to infer operational insights about the device.[22–26] In such applications, DD modelling combined with Bayesian inference has enabled estimation of mobile ion concentrations[27] while decision tree models have identified dominant recombination mechanisms.[24] Recently Kirchartz and Das[28] comprehensively reviewed the inverse problem across multiple experimental techniques, highlighting how Bayesian inference coupled with numerical modelling can extract meaningful information from experimental data to guide device optimization. Complementary approaches include the analysis by Kim et al.[29] analysis of charge transport in PSC from EIS data and Parikh et al.'s predictive model for Br-based PSC low-frequency response.[30] Nevertheless, to our knowledge, no comprehensive framework combining DD modelling with ML has been developed to systematically address the inverse problem of experimental impedance spectra with predictive reliability.

In this study, we selected a standard PSC configuration ($TiO_2$/MAPI/Spiro-OMeTAD) to evaluate the efficacy of ML in addressing the EIS inversion problem. For the DD simulations, we fixed readily measurable parameters, such as layer thicknesses, energy levels, band gaps, and optical/dielectric properties, while focusing on more elusive yet performance-critical physical parameters. These include bulk and interfacial recombination properties, as well as ionic characteristics (mobile ion concentration and mobility). A key aspect of our methodology was defining optimal 'features' for EIS data representation. We explored two approaches: (1) using an *instrumental* equivalent circuit model to extract circuit elements through fitting, and (2) directly identifying spectral signatures in EIS data (e.g., peak positions and amplitudes in frequency spectra). After feature extraction from DD-generated EIS datasets, we trained Random Forest (RF), deep neural network (DNN), and Gradient Boosting Regressor (GBR) models to assess the predictability of each target physical parameter. We further analyzed the influence of each feature on the predictions using the model-agnostic post-hoc explanation method Shapley Additive Explanations (SHAP)[31] on the best performing models. Finally, we fabricated experimental devices with identical configuration and measured their impedance responses under both short-circuit (SC) and open-circuit (OC) conditions. These experimental spectra served to validate the ML models' accuracy in predicting the extractable physical parameters.

This manuscript is organized as follows. First, we outline the methodology for generating DD simulation data, along with details of experimental procedures and ML analysis. Next, we present and discuss results demonstrating the models' predictive power for each target physical parameter. This section also includes model interpretability analysis and a comparison between predicted and experimental EIS spectra. Finally, we present key conclusions derived from our study.

# METHODS

The selected configuration on which this study is based is a standard $TiO_2$/MAPI/Spiro-OMeTAD configuration with fixed layer thicknesses under 1-sun illumination. In **Table 1**, details of the six physical parameters (Y values, targets) chosen for the development of the ML models are shown. **Figure 1** summarizes the general strategy followed to develop the models.

**Table 1.** Physical parameters (targets) studied and sweep intervals. Note: in the models, these parameters entered in a logarithmic scale.

| Parameter code | Parameter name | Lowest value | Highest value | Units |
|---|---|---|---|---|
| A | Bulk electron pseudolifetime | $10^{-7}$ | $10^{-5}$ | s |
| B | Bulk hole pseudolifetime | $10^{-7}$ | $10^{-5}$ | s |
| C | Mobile ion density | $10^{17}$ | $10^{19}$ | $cm^{-3}$ |
| D | Ion mobility | $10^{-12}$ | $10^{-10}$ | $cm^2V^{-1}s^{-1}$ |
| E | Surface recombination velocity of holes across $TiO_2$ interface | 0.5 | 50 | m/s |
| F | Surface recombination velocity of electrons across spiro interface | 0.5 | 50 | m/s |

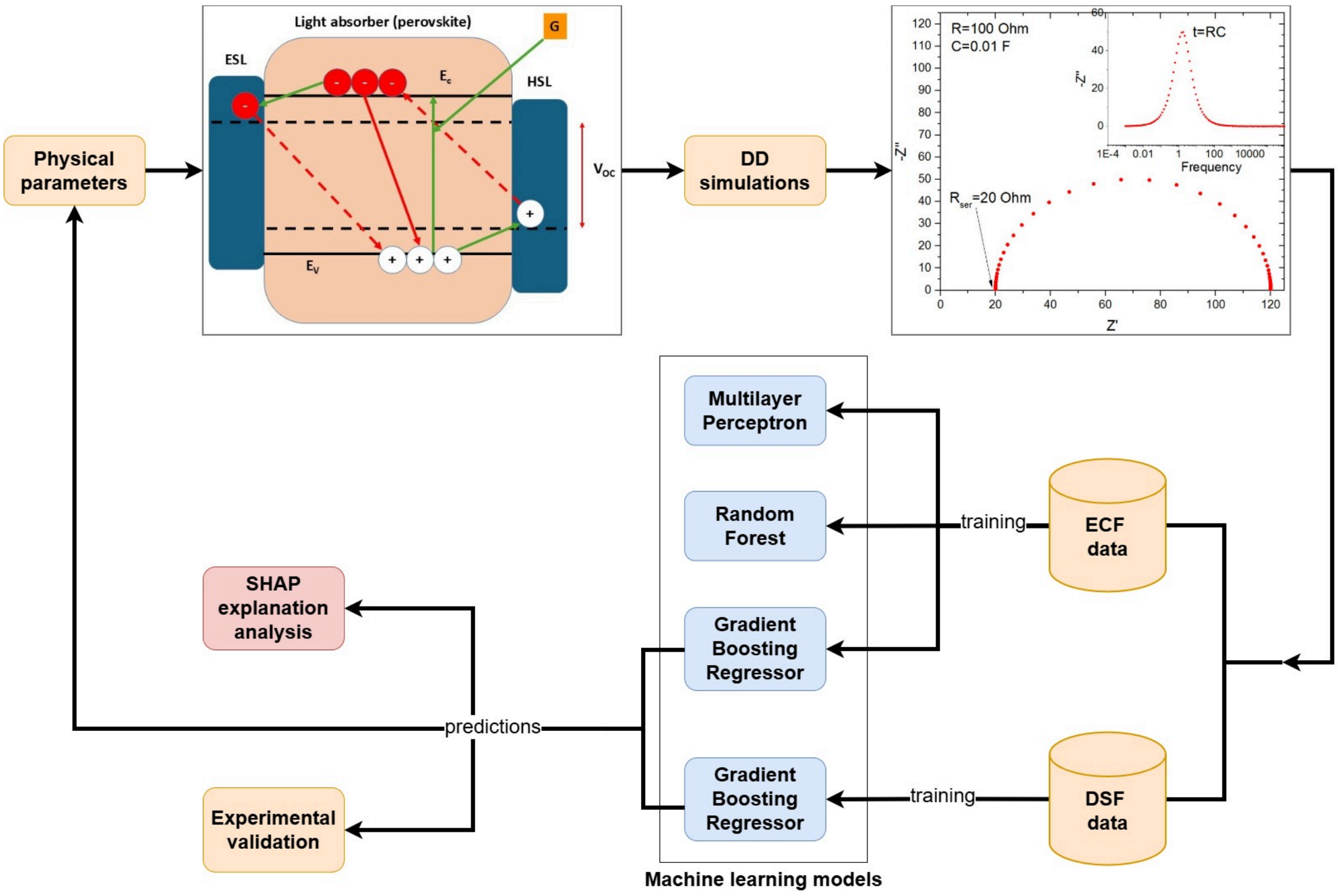

**Figure 1**. Overview of the workflow used for data generation, model training, and validation. Physical parameters are first sampled and passed to DD simulations to produce two distinct feature sets: Equivalent Circuit Fitting (ECF) and Direct Signal Featurization (DSF). These simulated datasets are then used to train multiple machine learning models to learn the inverse

mapping from features to physical parameters. The trained models generate predictions, which are subsequently analyzed through SHAP-based explainability and validated against experimental measurements.

## Drift-diffusion modelling

Extensive DD simulations for a wide range of the A to F parameters were carried out under 1 sun illumination and OC and SC conditions using SETFOS v5.5 software (Fluxim inc.)[32] This software solves numerically transport (drift + diffusion) equations for electrons, holes and ions in a stack of optically and electronically active materials. The numerical problem includes photogeneration (for a given illumination, AM1.5G in this case) and recombination (both in the bulk and at the interfaces).

In the simulations, a random sweep of A to F values in the intervals indicated in Table 1 was implemented. Parameters A and B refer to recombination of free carriers in the bulk of the perovskite, whereas E and F quantify the recombination velocity of the respective minority carrier at the electron-selective ($TiO_2$) and hole-selective (Spiro) interface respectively. In addition, C and D represented the ionic properties of the perovskite, concentration and mobility of positively charged ions respectively. Importantly, the sweep ranges for these parameters were deliberately constrained to physically motivated and literature-supported domains representative of MAPI-based perovskite solar cells, as established by a broad body of experimental and modelling studies.[5,13,16,17,21,33–39]

A preliminary sweep in a regular grid served to analyse the relative impact of each physical parameter on the extracted features prior to ML analysis (see Supporting Information **Figures S2** and **S3**). This preliminary calculation showed that surface recombination of the majority carrier does not influence the results and was, consequently, discarded in the subsequent random sweeps used for ML training.

Two sets of spectra were generated via DD simulations: one with 235 spectra (see distribution in **Figure S4**), which was used for manual equivalent circuit fitting, and one with 4000 spectra, which was used for two different approaches to automatic featurization, as described below.

In **Figure 2** and in the Supporting Information details of the methodology and the configuration considered in the simulations are collected. The fixed parameters of the simulations are shown in **Table S1**. Impedance spectra were generated between 0.1 Hz and 1 MHz. This frequency window is widely adopted in impedance spectroscopy studies of perovskite solar cells, including MAPI-based devices, and is sufficient to resolve both high-frequency recombination-related processes and the dominant low-frequency response associated with ionic and interfacial effects.[5,9,40,41]

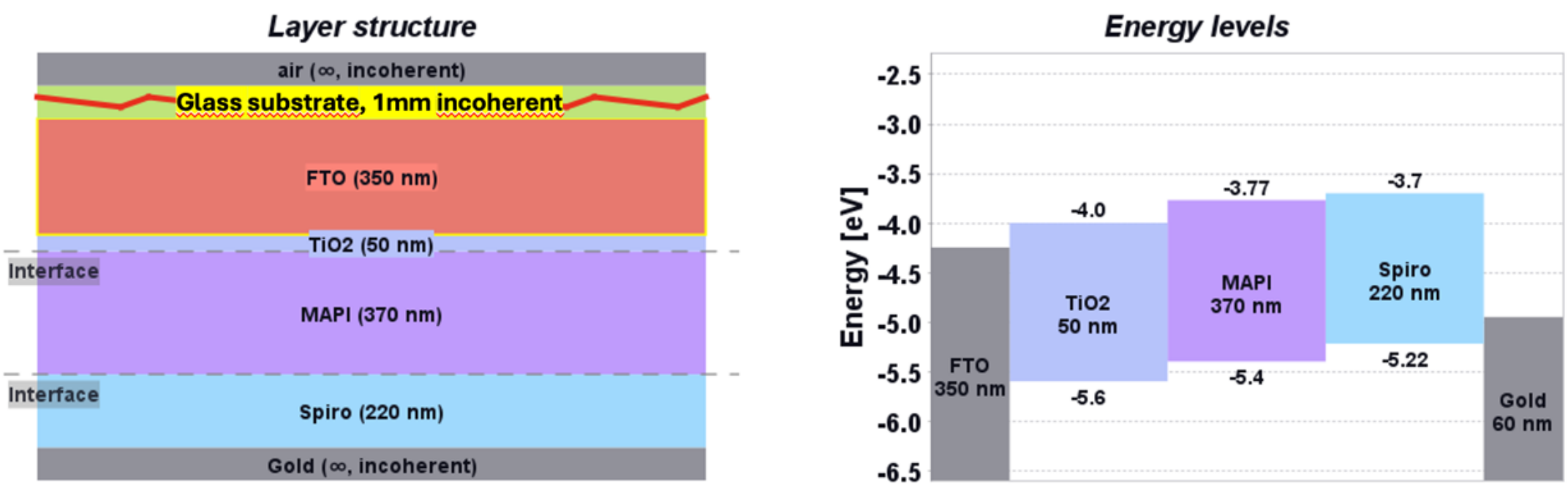

**Figure 2**. Device stack and energy-level alignment used in the drift–diffusion/EIS simulations. (a) Layer structure of the planar n–i–p cell: Glass substrate (1 mm) / FTO (350 nm) / compact $TiO_2$ (50 nm) / MAPI (370 nm) / Spiro-OMeTAD (220 nm) / Au; air (top) and Au (bottom) are treated as optically incoherent semi-infinite media. The two dashed "Interface" markers indicate where interfacial recombination E ($TiO_2$/MAPI) and F (MAPI/Spiro) are defined (ranges in Table 1). (b) Vacuum-referenced energy-level diagram (eV) used in the simulations.

## Spectra featurization and machine learning models

Two strategies were followed to obtain the features describing the instances used for training the ML models (feature coding is presented in **Table S2**). First, with **equivalent circuit fitting (ECF)**, the set of 235 DD-simulated and the experimental spectra were individually fitted to an equivalent circuit model consisting of a resistance connected in series with up to four RC elements. Zview software (Scribner) was used to carry out the fits. The RC elements were allowed to exhibit non-ideal capacitor behaviour, and to adopt negative values when the signal appeared in the positive imaginary quadrant. The first, second, and third RC elements were assigned to represent the high-, intermediate-, and low-frequency responses of the impedance spectra, respectively. A fourth RC element was included only when the low-frequency signal exhibited a change of sign in the imaginary component (i.e., when the response initially appeared in the positive imaginary quadrant and then returned to the negative one). This behaviour was observed in some spectra under OC conditions but not under SC ones. These fits produced a set of up to 12 equivalent circuit features characterizing each spectrum (excluding the series resistance).

In order to create a larger dataset with ECF features, we carried out automatic fitting of spectra (using an adaptation of the *impedance.py* code[42]). This allowed the batch fitting of 4000 spectra, avoiding the need for manual fitting of individual spectra in Zview. The challenge of automatic ECF is that sensible initial circuit parameters must be provided; therefore our strategy consisted of trying all the 235 circuit parameter combinations obtained in the manual fitting step as candidate initial conditions in each fitting, and retaining only the fitting that led to the lowest chi-squared for each of the 4000 spectra. This procedure is justified by the fact that the 4000-sized dataset is generated from physical parameters in the same ranges as those used for the 235-sized dataset, so each spectrum in the larger set is close to one of the spectra in the smaller set, as illustrated in **Figure S5**.

Second, with **direct signal featurization (DSF)**, each EIS instance was numerically processed to detect peaks in the imaginary part of the impedance. For each peak, the frequency position, the height and the full width at half maximum (FWHM) were stored as features. Additionally, the low frequency limits (0.1 Hz) of the real and the imaginary part of the impedance are also recorded and stored as extra features.

Although both strategies led to some similar conclusions (see below), it is convenient to discuss the pros and cons of each. ECF could have the advantage that each circuit element might be given a physical interpretation in the form of a resistive or a capacitive process. However, it is important to note at this point that the equivalent circuit chosen was not strictly aimed to provide physical insights. ECF-based analysis serves two roles: (1) to provide a compact and interpretable representation of impedance spectra in terms of standard EIS descriptors (resistances, capacitances, and time constants) that are familiar to the impedance community, and (2) to offer a qualitative bridge between conventional impedance analysis and the DD-based ML inversion. ECF is not used to extract physical parameters, but to verify that the ML model is learning from recognizable spectral features rather than arbitrary numerical patterns. Hence, the role of the equivalent circuit is just to enable a feature extraction procedure for defining each EIS instance and obtaining raw numerical data for the learning process.

However, the ECF strategy has an important drawback: the numerical fittings are not easily automated and require human attention for each instance. In this work, we solved the problem by using the 235 individually (manually) fitted spectra to provide sensible initial conditions for the automatic fitting, but this strategy is laborious. As an additional simplification, although the series resistance was considered in the ECF, its values were not included in the final dataset. This is justified given the experimental origin of the series resistance, commonly associated to contacting and wiring, which are conditions not considered in the DD modelling. Hence, only 9 (12) features (3(4) resistances, 3(4) capacitances and 3(4) capacitor ideality factors) were used in the training process in the SC (OC) experiment. It must be noted that in some cases a very large value of the low-frequency resistance was found. To account for infinite resistances, we used a conductance parameter (G3) instead of a resistance (R3) as a feature in the third RC element.

In contrast, DSF allows for automated extraction of features for each EIS instance with minimal human labor. To keep the number of features of the same order as in the ECF, a peak height sensitivity was introduced in the automated extraction so that the number of peaks recorded was limited to two: high and low-frequency peaks, respectively. The number of peaks found by the DSF (either 1 or 2) was also included as a binary feature in the ML data set. Additionally, the low-frequency limits of the real and imaginary part of the impedance were recorded and included in the feature data set. As an additional correction, for those cases in which the high-frequency peak maximum lies beyond the frequency interval considered (0.1 -$10^6$ Hz), the peak maximum is assumed to be the imaginary value of the impedance at $10^6$ Hz).

From the ML point of view, the inverse problem involves predicting six numerical targets and thus constitutes a multi-output regression task. The selected models were Multilayer Perceptron (MLP), Random Forest (RF)[43] and Gradient Boosting Regressor (GBR).[44] The MLP is a feed-forward neural network using two fully connected layers. RF and GBR are both ensemble learning methods that combine multiple decision trees to enhance predictive accuracy. RF generates multiple trees on random subsets of the data and averages their predictions, while GBR builds trees sequentially, with each tree correcting the errors of its predecessor. For all models, nested cross-validation and hyperparameter tuning were implemented to identify the configuration with maximum predictive power without overfitting. For the MLP, the hyperparameters tested include different configurations of the size of layers (50 and 100 neurons), ReLU and hyperbolic tangent as transfer functions, Adam[45] or stochastic gradient descent as optimizers, and a learning rate of 0.0001, 0.001, or 0.01. For GBR, the hyperparameters tuned included the number of estimators (200, 600), learning rate (0.01, 0.1), maximum depth (5, 10), minimum samples per split (5, 10), and subsample ratio (0.8, 1.0). For RF, the number of estimators (100, 200, 500), maximum depth (None, 10, 20), and minimum samples per split (2, 5, 10) were explored. MLP and RF natively support multi-output regression, allowing it to optimize all targets simultaneously. In contrast, GBR implementation fits a separate model for each target independently. This difference implies that RF can directly minimize a joint error across all outputs, whereas the GBR minimizes the error for each target individually, which may influence the overall predictive performance. However, as shown below, GBR achieves reliable predictions of all physical parameters that contribute to the overall impedance response. Still, inter-parameter correlations are not explicitly considered. Although very interesting for the understanding of the impedance, this further line of research lies beyond the scope of the present study.

In the case of the 4000-sized dataset obtained with automatic ECF, we only used the best-performing GBR model. In the automatic ECF approach, we had to deal with the varying quality of fitting to equivalent circuits, which is characterized by a chi-squared value for each instance. During training each instance was given a weight proportional to exp{-log(chi-squared)/$T$}, where $T$ is a fictitious temperature that characterizes the down-weighting of the poorer-fitted spectra, and it was treated as a hyperparameter. The automatic ECF worked less well for the SC than for the OC spectra, so for SC we had to remove (rather than just down-weight) very poorly fitted spectra completely from the training set, to avoid introducing noise; the threshold chi-squared was also treated as a hyperparameter. All models were implemented using the Python *scikit-learn* library.

## Experimental

In order to validate the developed ML models, $TiO_2$/MAPI/Spiro-OMeTAD test devices were fabricated and characterized via measurement of current-voltage (IV) curves and impedance experiments at OC and SC conditions under 1-sun simulated solar AM1.5G illumination. Experimental PCE values of 15 devices range between 15.3 % and 16.5% (mean 15.9%), which are in the order of the state of the art for PSC devices with MAPI as absorber. Details of the fabrication procedures and the experimental analysis are reported in the Supporting Information file.

## RESULTS AND DISCUSSION

**Table 2** presents a summary of the performance achieved by each model and data configuration, for each of the targets in terms of $R^2$ score. The $R^2$ score is defined as the proportion of variance in the target variable explained by the model relative to a baseline predictor that always outputs the mean. In particular, the prediction results for the original 235-sized ECF test set in the OC experiment are presented in **Figure S6** (MLP), **Figure S7** (RF), and **Figure S8** (GBR). The comparison of the three models shows the superior performance of the GBR model in predicting the physical parameters from the impedance spectra.

**Table 2.** Summary of ML results and $R^2$ score for the six physical parameters (targets).

| **Experiment** | **Feature extraction method** | **Number of instances** | **Model** | **A** | **B** | **C** | **D** | **E** | **F** |
|---|---|---|---|---|---|---|---|---|---|
| OC | ECF | 235 | MLP | 0.47 | 0.41 | 0.67 | 0.31 | 0.33 | 0.36 |
| OC | ECF | 235 | RF | 0.69 | 0.66 | 0.73 | 0.43 | 0.59 | 0.37 |
| OC | ECF | 235 | GBR | 0.80 | 0.84 | 0.82 | 0.68 | 0.63 | 0.50 |
| SC | ECF | 235 | GBR | 0.73 | 0.69 | 0.86 | 0.86 | -0.24 | 0.94 |
| OC | ECF (aut.) | 4000 | GBR | **0.94** | **0.91** | **0.92** | 0.85 | 0.81 | 0.77 |
| SC | ECF (aut.) | 4000 | GBR | 0.84 | 0.89 | **0.90** | **0.96** | -0.09 | **0.97** |
| OC | DSF | 4000 | GBR | 0.16 | 0.14 | 0.43 | 0.66 | 0.64 | 0.49 |
| SC | DSF | 4000 | GBR | 0.54 | 0.46 | 0.47 | 0.88 | -0.06 | 0.85 |

Notes: A = Bulk electron pseudolifetime; B = Bulk hole pseudolifetime; C = Mobile ion density; D = Ion mobility, E = Surface recombination velocity of holes across $TiO_2$ interface; F = Surface recombination velocity of electrons across spiro interface

In **Figure 3**, the GBR model is applied to the automated 4000-sized set. In general, the predictions are remarkably accurate with $R^2$ score values over 0.9 for the bulk recombination parameters A and B. In contrast, surface recombination (parameters E and F) is not predicted as accurately. Ionic parameters also showed a good $R^2$ score between ground truth and predicted values, especially for the ion concentration C ($R^2$ score = 0.93). In general, impedance data obtained under OC conditions proved to be sufficient for accurately predicting recombination and ionic parameters.

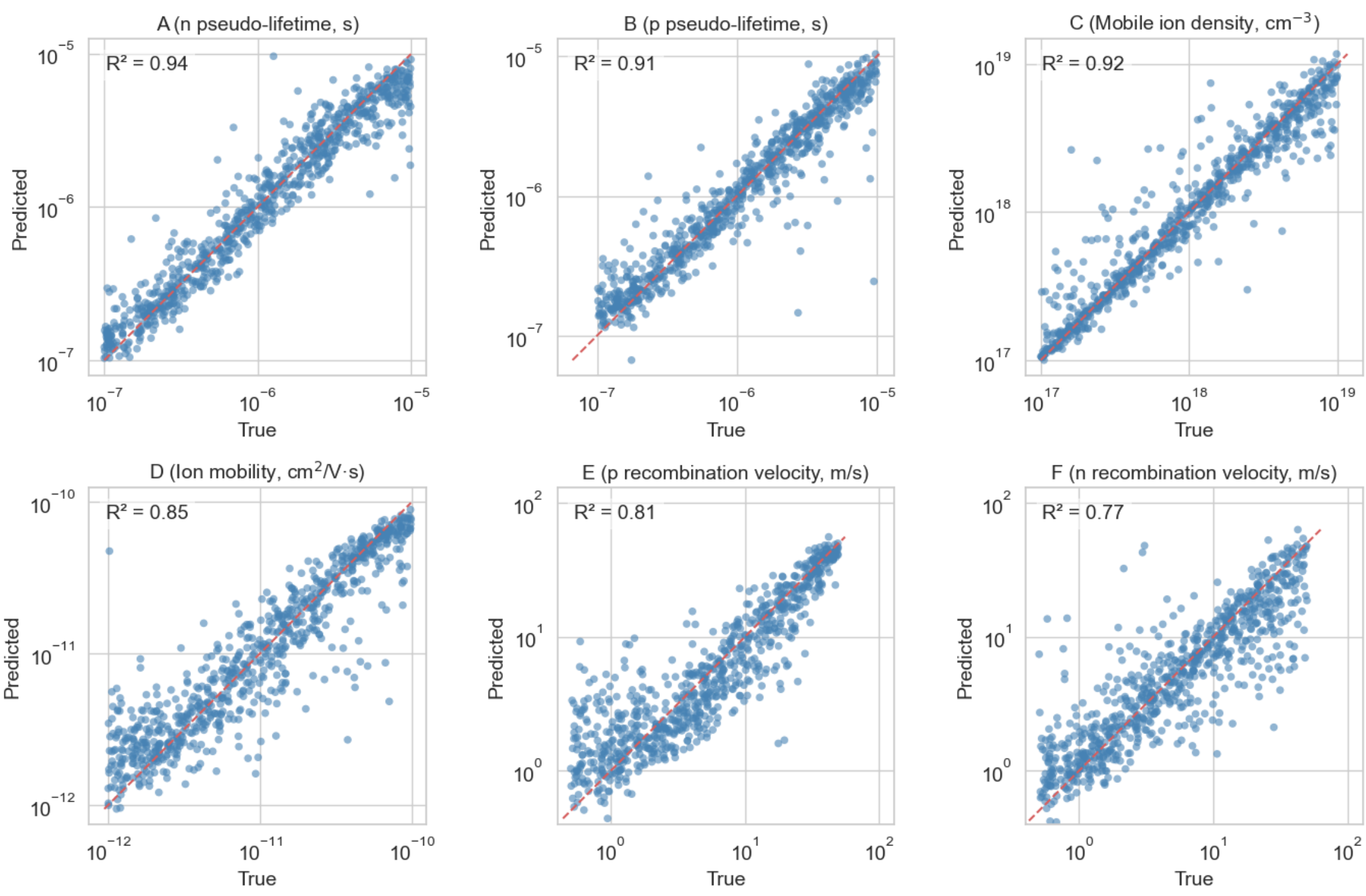

**Figure 3.** Predicted versus true values in test set for each target variable for the GBR model using automatic ECF featurization under OC conditions.

An important question is whether EIS measurements under short-circuit (SC) conditions might also provide accessible physical insights into device operation. Results from the GBR model are presented in **Figure S9** (original 235-sized ECF set) and **Figure 4.**

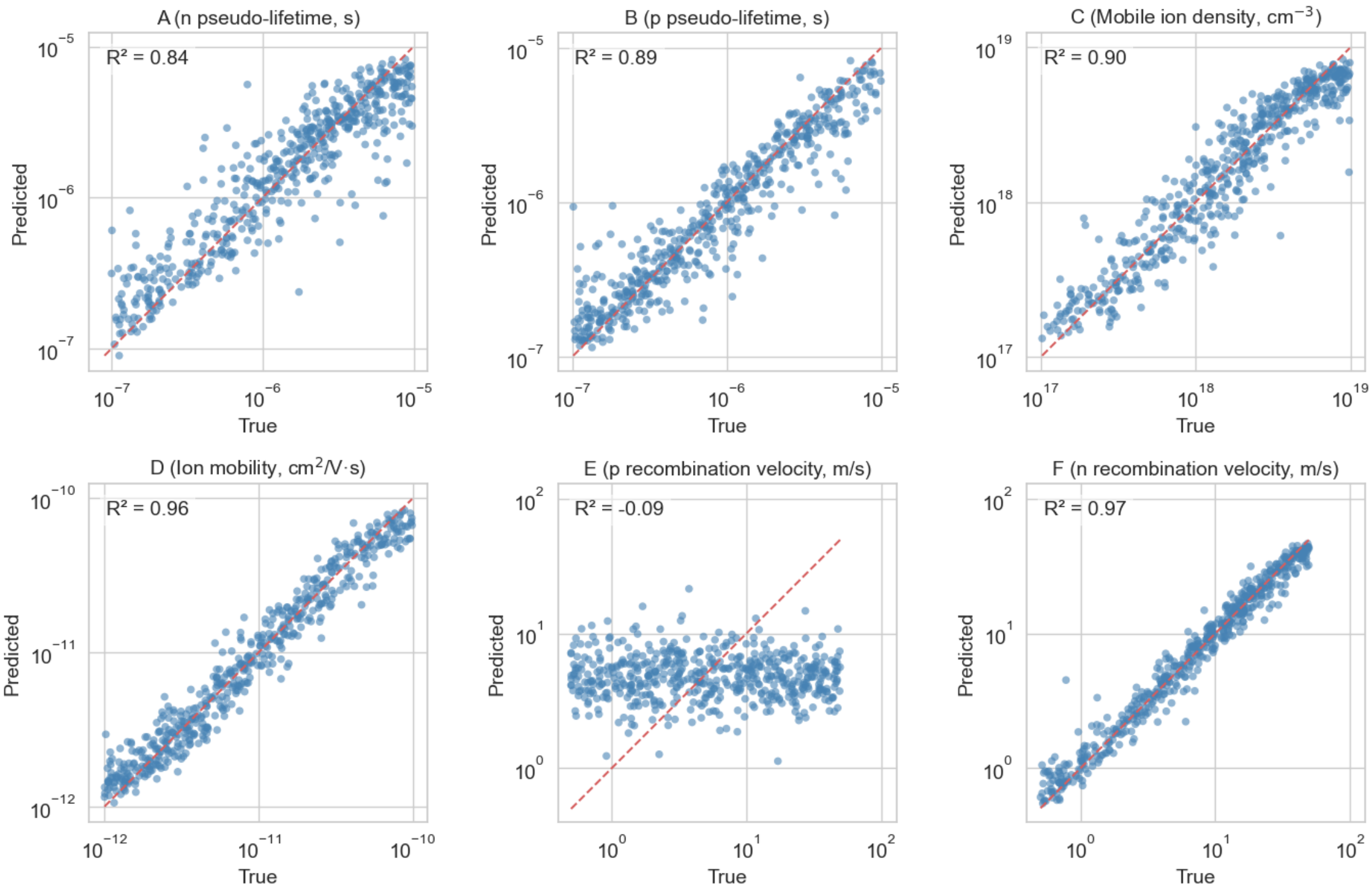

**Figure 4.** Predicted versus true values in test set for each target variable for the GBR model using automatic ECF featurization under SC conditions.

The SC condition analysis revealed that ionic and bulk recombination properties remain predictable. However, the $R^2$ score for the A and B parameters is lower than under OC conditions, whereas the prediction of the ion mobility (parameter D) is greatly improved. Interestingly, no variance ($R^2$ score close to zero) is predictable for the hole recombination at the electron selective layer (parameter E), while the prediction of its symmetric counterpart (parameter F) is exceptionally good. The apparent anticorrelation of parameter E does not represent a physical effect (the small negative value is meaningless), but instead reflects the fact that recombination at this interface becomes effectively invisible to impedance spectroscopy under short-circuit conditions as discussed below.

The ML analysis of the impedance data demonstrates that OC measurements are preferable when investigating recombination losses, while SC conditions may be more suitable for studying charge transport properties. This distinction arises because the higher stored charge density under OC conditions amplifies recombination signatures (charge-density-dependent processes) whereas SC conditions naturally emphasize transport phenomena due to the established current flow.

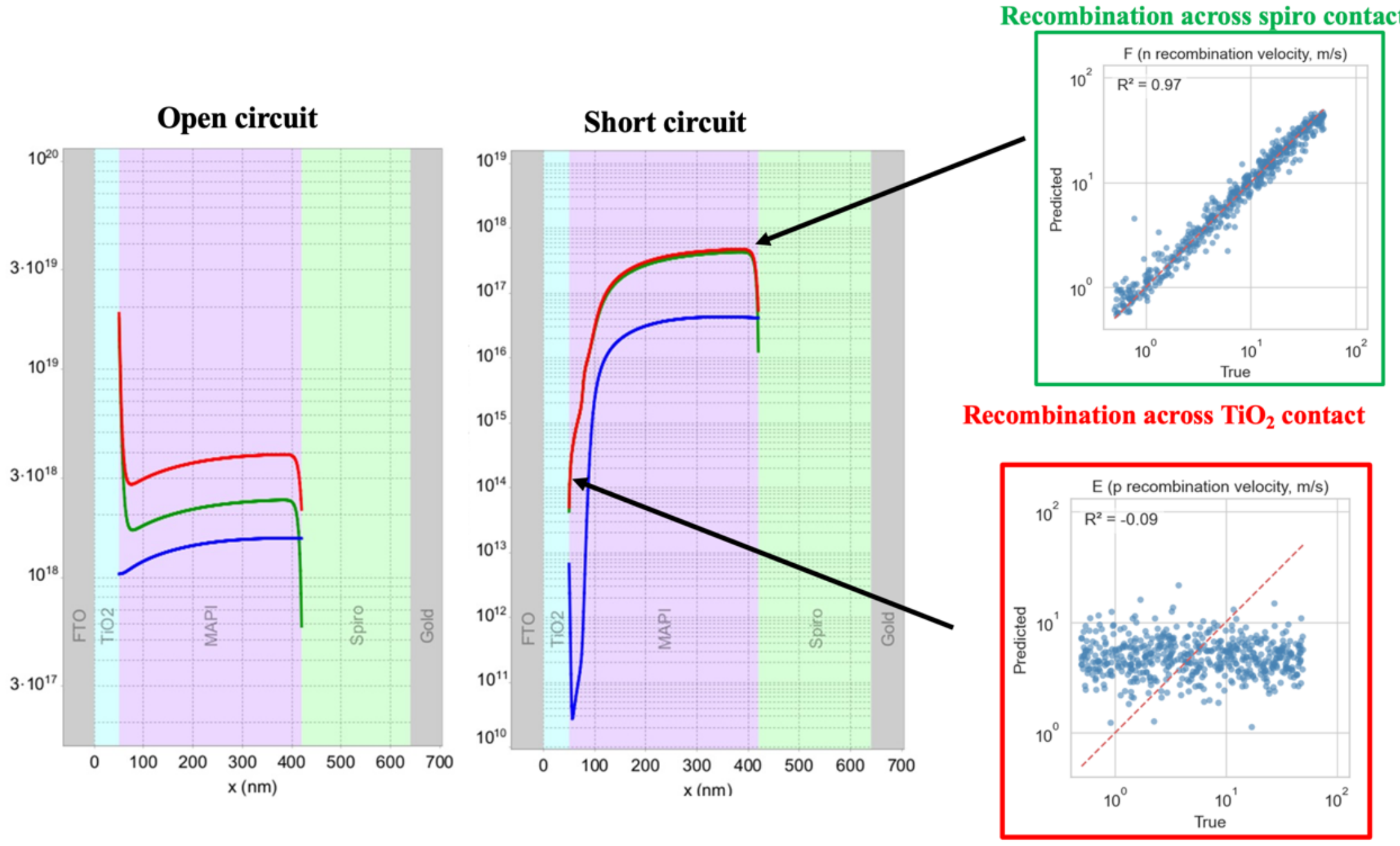

**Figure 5.** Illustration of the performance the ML analysis for the prediction of surface recombination velocities. Left: recombination rate profiles in $cm^{-3}$ $s^{-1}$ under open and short-circuit conditions as obtained by DD modelling (red: total recombination, green: trap mediated recombination, blue: bimolecular recombination). Right: predicted versus true values for parameters E (recombination through $TiO_2$ contact) and F (recombination through the spiro contact) for the GBR model using ECF featurization under SC conditions.

It is interesting to determine the origin of the asymmetry in the predictions of the surface recombination parameters E and F. In **Figure 5** and **S10** we focus on surface recombination only and include the recombination rate profiles as obtained by the numerical solutions of the DD equation under both OC and SC conditions. As expected, the gradient in both types of profiles is stronger under SC conditions. Significantly, the concentration of holes is particularly low in the vicinity of the $TiO_2$ interface in the SC case. Therefore, the recombination rate drops drastically (by four orders of magnitude) at this interface when the cell is kept under SC conditions. This observation explains why the E parameter cannot be predicted in an SC experiment, in contrast to parameter F, which shows a very high $R^2$ score of 0.97 (Figure 4). Thus, the gradients induced by the DC current generate an asymmetry in the recombination rates that critically determines the type of information that can be extracted from the impedance spectrum measured under SC conditions. The comparison between OC and SC results hence demonstrates that the predictability of a particular parameter, surface recombination in this case, is directly linked with the relative impact of that parameter in the generation of the impedance response.

To provide a visual idea of the prediction quality of the models, we selected several additional random instances not included in the original training set to compare the predicted spectra with the true ones (see **Figure S11**). The quality of the prediction depends on the instance picked but in general the ML reproduces reasonably well the shape of the spectrum and the size and position of the arcs and frequency peaks.

We further analyze the best-performing ML models of each configuration with the model-agnostic post-hoc explanation SHAP. This approach approximates Shapley values to estimate the contribution of each feature to a particular prediction. Positive SHAP values increase the prediction of the target feature with respect to the expected value for the dataset, while negative SHAP values indicate that the feature decreases the prediction. By aggregating individual feature importance values for each instance (local explanations), a global view of feature importance for each target can be obtained (global explanation). In **Figure 6,** a summary of the SHAP feature importance values for the target parameters A, C, D, and E  is shown. Results for the SC case are presented in **Figure S12.**

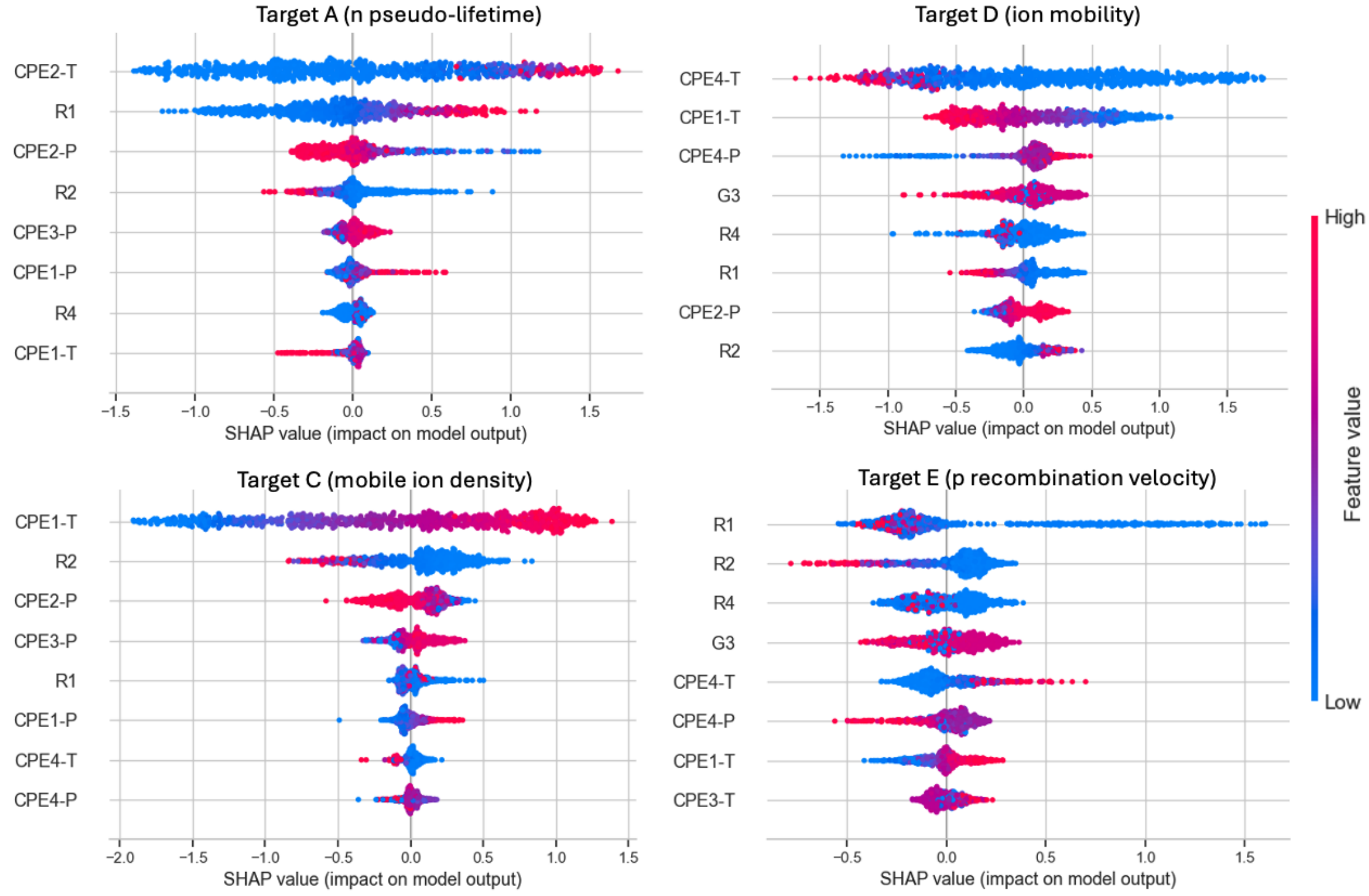

**Figure 6.** SHAP analysis for the GBR results under OC conditions with ECF featurization. SHAP analysis is interpreted in the following way: features are listed in order of importance. The SHAP

value shows the impact of that feature on the prediction of each target. A positive value pushes the prediction higher, a negative value pushes the prediction lower. Red colour means high feature value for that instance. Blue colour means low feature value for that instance. Feature codes: R*x*'s = resistances; CPE*x*-T's = capacitances, CPE*x*-P's = capacitor ideality factors, where $x = 1,2,3,4$ is counter running from high to low frequencies. G3= low frequency conductance.

The main findings from the SHAP analysis are the following: recombination is mainly "seen" in the high frequency resistance (R1), in such a way that a longer bulk carrier lifetime (parameters A and B) or a faster surface recombination velocity (parameters E and F) moves the resistance to larger and lower values, respectively. The mid-frequency capacitance also shows a strong impact on the bulk recombination loss. On the other hand, the ion density (parameter C) visibly correlates with the high-frequency capacitance. The high frequency capacitance has been associated to the geometrical capacitance of the device, which depends on the dielectric properties of the perovskite.[7,9,46] Our results prove that the concentration of mobile ions is a major factor affecting the dielectric response and the dielectric constant of the perovskite material. The ion mobility (parameter D) is mainly impacting the low frequency features of the spectrum. This is also consistent with previously reported results in the literature following DD analysis of impedance spectra in PSCs.[16,19] However, under SC conditions, it is found that the ionic mobility is also affecting the high-frequency resistance. In general, it is found that recombination influences the entire impedance spectrum rather than specific frequency components, which is a defining characteristic of perovskites, the coupling between fast and slow processes in the spectrum.[2] The results for SC are mainly consistent with those for OC for those parameters that can be predicted at SC.

As mentioned above, ECF is limited by the manual fits required to define the features for each spectrum (even in the case of automatic circuit fitting, the initial manual fitting was required to provide a sensible set of initial conditions). To explore a fully automated procedure for featurization DSF was implemented. This involves the batch-processed extraction of signal features from the spectra without the need for a manual fit to an equivalent circuit. In **Figure S13** (OC) and **Figure S14** (SC) results from the implementation of the GBR model are shown.

It is observed that, in general, ML predictions based on DSF provide consistent results with those obtained using ECF except for bulk recombination parameters (A and B), which showed worse correlation between true and predicted values under OC conditions but a better $R^2$ score under SC conditions. Again, the SC experiment proves to be more informative towards the extraction of transport parameters such as the ion mobility (parameter D) or the surface recombination velocity across the spiro interface (parameter F), with $R^2$ scores of 0.88 and 0.85, respectively. Overall, the $R^2$ scores for all targets are poorer using features obtained by DSF with respect to results obtained via ECF analysis. This clearly indicates that an analysis based on detection of peak signals (position, height and FWHM) plus low- and high-frequency limits of the impedance is not sufficient to capture all the information that can potentially be extracted from an impedance experiment. This observation is especially revealing for the case of bulk recombination, which is, as a matter of fact, neither affecting the low frequency part of the spectrum or the peaks detected in the imaginary part of the impedance. Thus, a more extensive analysis, either by detailed equivalent circuit fitting or using the full spectral data would be more adequate to obtain physical parameters from impedance measurements.

Summaries of the SHAP analysis can be found in **Figure 7** (OC) and **S15** (SC). The main observation is that ionic parameters (C and D) are mostly featured in the low frequency limit of the real part of the impedance. The ionic density is also found to affect the frequency positions of the peaks, especially at low frequencies. In general, the widths of the signals in the imaginary part of the spectrum do not appear to appreciably impact the predictions of any of the targets. This indicates that the peak shape in the imaginary part of the impedance feature does not bear significant information on the physical parameters. On the other hand, the surface recombination velocity (target F) is found to significantly determine the position of the high frequency peak, both under OC and SC conditions, a result which is in line what the results obtained using ECF analysis for the high frequency resistance. Thus, a faster recombination is correlated with a higher value of the position of the high-frequency peak, a well-known result in solar cells modelling.[3,7]

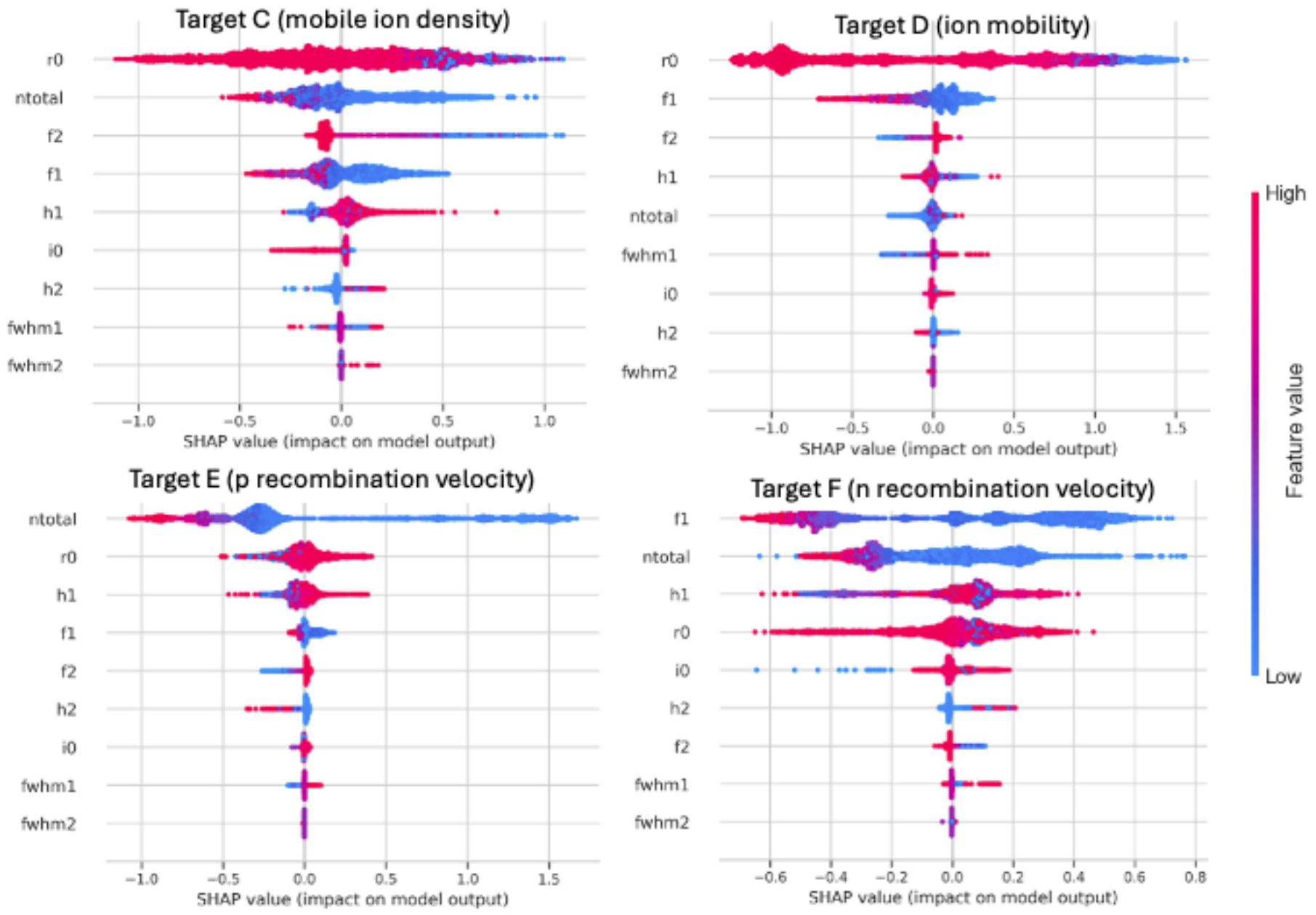

**Figure 7.** SHAP analysis for the GBR results under OC conditions with DSF featurization. Feature codes: f*x*'s = peaks frequency positions; h*x*'s = peaks heights, fwhm*x*'s = peaks widths, where *x* = 1,2 is counter running from high to low frequencies. r0, i0, low frequency limits of the real and the imaginary impedance, respectively. Ntotal = number of peaks found.

Following the ML analysis, we next examine the inverse problem, that is, assessing whether the trained models can accurately predict the true values of the target physical parameters. To do this, both ECF and DSF types of feature extraction were applied to experimental EIS data for the $TiO_2$/MAPI/spiro configuration under 1 sun illumination and the extracted features were input to the ML trained models to get predictions of the physical parameters A to F. We focus our study on the GBR model, which yielded the best performance in terms $R^2$ score between true and predicted values for all targets as shown in Table 2.

In **Table 3** results for the predicted values of the targets are collected. Note that additional "mixed" predictions were also generated by combining the best predictions of the simulations under SC and OC conditions which, as discussed above, show different predicting power depending on the target. In **Figure 8** a comparison of measured spectra under both OC (1V) and SC (0V) conditions and the results of DD simulations for the predicted targets is presented. For clarity, only the results of the DD simulations for set 3 and set 6 are shown. All predictions are collected in **Figure S16.**

**Table 3.** Prediction of the GBR models from experimental EIS data. * = best predicted from 1, best predicted from 2. ** = best predicted from 4, best predicted from 5. *** = best predicted from 7, best predicted from 8. The most reliable predictions are given in bold font, with the ±1 prediction interval (the intervals are asymmetric because they are based on the RMSE in log scale)..

| **Simulation** | **Experiment /Featurization** | **A (s)** | **B (s)** | **C ($cm^{-3}$)** | **D** ($cm^2V^{-1}s^{-1}$) | **E (m/s)** | **F (m/s)** |
|---|---|---|---|---|---|---|---|
| **1** | **OC/ECF** | $1.6\times10^{-7}$ | $5.7\times10^{-7}$ | $1.4\times10^{17}$ | $6.8\times10^{-11}$ | 8.97 | 23.6 |
| **2** | **SC/ECF** | $1.0\times 10^{-6}$ | $5.0\times10^{-7}$ | $3.3\times10^{17}$ | $5.0\times10^{-11}$ | 7.50 | 35.2 |
| **3*** | **Mixed 1 & 2** | $1.6\times 10^{-7}$ | $5.7\times 10^{-7}$ | $3.3\times10^{17}$ | $5.0\times10^{-11}$ | 8.97 | 35.2 |
| **4** | **OC/ESF (aut.)** | **$1.3\times 10^{-6}$** | **$4.52\times 10^{-6}$** | **$1.109\times 10^{17}$** | $4.44\times 10^{-12}$ | **10.5** | 25.3 |
| **5** | **SC/ESF (aut.)** | $6.3\times 10^{-7}$ | $7.5\times 10^{-7}$ | $3.5\times 10^{17}$ | **$7.1\times 10^{-11}$** | **6.8** | **41.2** |
| **6**** | **Mixed 4 & 5** | **$1.3\times 10^{-6}$ ($0.92\times10^{-6}$ – $1.8\times10^{-6}$)** | **$4.5\times 10^{-6}$ ($3.1\times10^{-6}$ – $6.7\times10^{-6}$)** | **$1.1\times 10^{17}$ ($0.74\times10^{17}$ – $1.6\times10^{17}$)** | **$7.1\times 10^{-11}$ ($5.4\times10^{-11}$ – $9.2\times10^{-11}$)** | **10.5 (5.8 – 18.8)** | **41.2 (33.0 – 51.5)** |
| 7 | **OC/DSF** | $9.3\times10^{-7}$ | $1.0\times10^{-6}$ | $3.5\times10^{18}$ | $3.9\times10^{-11}$ | 19.4 | 9.3 |
| 8 | **SC/DSF** | $5.8\times10^{-7}$ | $5.3\times10^{-7}$ | $4.3\times10^{18}$ | $1.3\times10^{-12}$ | 2.72 | 33.9 |
| **9***** | **Mixed 7 & 8** | $5.8\times10^{-7}$ | $5.3\times10^{-7}$ | $4.3\times10^{18}$ | $1.3\times10^{-12}$ | 19.4 | 33.9 |

Notes: A = Bulk electron pseudolifetime; B = Bulk hole pseudolifetime; C = Mobile ion density; D = Ion mobility, E = Surface recombination velocity of holes across $TiO_2$ interface; F = Surface recombination velocity of electrons across spiro interface

Focusing on the ECF feature extraction method (simulations 1 to 6), the application of the GBR model to the two experimental measurements shows, in general, good consistency in the predictions of the physical parameters when we compare what we obtain from the OC data with what we obtain from the SC data. This is an encouraging result, as we can expect that the true physical parameters should be independent from the conditions of the experiment. The predictions of the ECF analysis with automatic fitting yield basically the same outcome. Thus, ion concentrations between $1.3\times10^{17}$ and $3.5\times10^{17}$ $cm^3$ and ion mobilities between $5\times10^{-11}$ and $7.1\times10^{-11}$ $cm^2V^{-1}s^{-1}$ are predicted for the studied configuration from the models with best $R^2$ scores.

Surface recombination velocities of 7-9 and 23-41 m/s are also predicted for the E and F parameters, respectively. These values are in line with the experimentally and physically accepted range for perovskite contact recombination.[47–49] and therefore represent physically plausible interface recombination velocities for the device architectures investigated in this work. In contrast, predictions for the bulk carrier lifetimes (parameters A and B) are not very reliable as can be inferred from the wider range of predicted values. When looking at the DSF results (simulations 4 to 7), the predictions for D and F were in line with the ECF ones, although the values obtained for the ion concentrations (parameter C) lay one order of magnitude above. However, considering the lower $R^2$ score found for this parameter (see Figures S13-14), it is reasonable to expect that the real value of the ion concentration is closer to values around $10^{17}$ $cm^3$. This value compares well with the prediction of McCallum et al.[27] using Bayesian optimization.

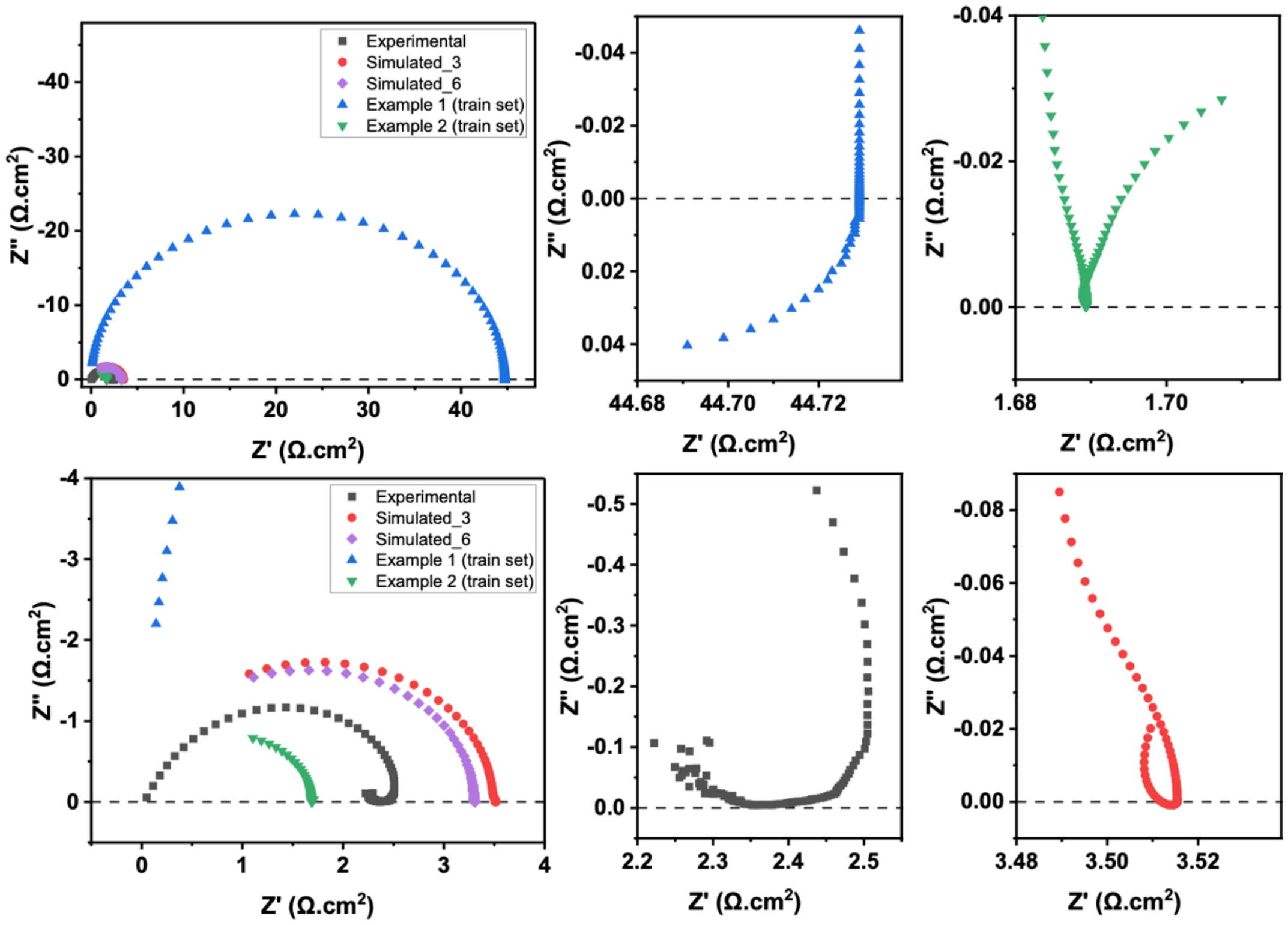

**Figure 8.** Comparison between experimental and simulated EIS spectra under OC conditions using ML-predicted spectra by DD simulation (parameters set 3 and set 6 in Table 3). The figure includes zooms of the spectra at low frequencies.

Comparison of the simulated spectra with the experimental ones (Figure 8 and S16) illustrates the difficulty to capture their full shape and size. To show the sensibility of the spectra to the choice of physical parameters we added to the plots simulated spectra for extreme values of the features considered in the train set. As it is well known, the size of the main arc of the spectrum is connected to the R1 feature. Considering the strong variability of the train set, the predicted spectrum lies relatively closed to the experimental one. Interestingly enough, it is observed that the prediction also nicely captures the shape of the loop found in the spectrum in the low frequency part.

## CONCLUSIONS

In this work, the intricacy of the impedance response of perovskite solar cells has been addressed using Machine Learning methods and the simulation capabilities of drift-diffusion numerical modelling. ML analysis as tested for a standard perovskite solar cell configuration provided two types of information.

Firstly, we established which physical parameters can be safely derived from the spectrum depending on the condition of the experiment: either at open circuit or at short circuit. We concluded that open-circuit conditions are especially suited to obtain recombination parameters whereas short-circuit conditions are recommended for the determination of ionic parameter such as mobilities. We also found relevant correlations between the target physical parameters and the features used to characterize the spectrum. In this regard we found that a method based on fittings to a generalized equivalent circuit contains more information and have stronger predicting power towards ML analysis than a simple direct collection of signals in the spectrum. In general, recombination, both in the bulk and at the interfaces, is detected in the high-frequency features, whereas ion mobilities tend to affect more the low-frequency part. However, the ion density is found to affect the high-frequency capacitance as well. The wide dependencies revealed between features in the spectra and physical parameters emphasize the difficulty to interpret the impedance response in perovskite solar cells.

On the other hand, the best ML model trained and tested (gradient boosting regressor) on actual experimental data predicted ion concentrations between 1.1 and 3.3 $10^{17}$ $cm^3$ and ion mobilities between 5 and 7 $10^{-11}$ $cm^2V^{-1}s^{-1}$ for a MAPI solar cell. Surface recombination velocities of 7-9 and 23-40 m/s are also estimated for the same configuration. The predicted spectra are found to reproduce the shape and fine details of the experimental ones, especially in consideration of the wide and strong sensibility of the impedance data on the choice of physical parameters.

In summary, the ML approach proves to be a powerful tool to address the inverse problem in the impedance analysis of complex systems such as perovskite solar cells. This is based on a physically grounded, system-specific framework that links impedance measurements to transport, recombination, and ionic parameters within a well-defined perovskite device architecture. This provides a practical and interpretable route for extracting physically relevant information from EIS, knowing what parameters can be safely extracted and which ones not.

## Acknowledgment

We acknowledge the Ministerio de Ciencia e Innovación of Spain, Agencia Estatal de Investigación (AEI) and EU (FEDER) under grants PID2022-140061OB-I00 (DEEPMATSOLAR) and PCI2024-153456 (INDYE). RGC and IGG acknowledge travel funding from the ENFIELD project ((European Lighthouse to Manifest Trustworthy and Green AI).)

# Inversion of the impedance response towards physical parameter extraction using Machine Learning – Supporting Information

*Mahmoud Nabil, Isel Grau-García, Ricardo Grau-Crespo, Said Hamad and Juan A. Anta,**

## S1. Drift-diffusion simulations.

SETFOS v5.5 software (Fluxim inc.)[1] , that solves drift-diffusion equations for a give stack of optically and electronically active materials was used to generate simulated impedance spectra under 1-sun AM1.5G illumination under OC and SC conditions. The parameters used in the calculations are shown in **Table 1** (variable parameters, used to train the ML models) and **Table S1** (fixed parameters).

**Table S1.** Fixed device stack, material constants, and simulation settings used to generate the drift–diffusion EIS data (values swept in Table 1 are not repeated). Symbols/abbreviations not otherwise defined in the table: *n* = refractive index; *k* = extinction coefficient (both wavelength-dependent); DOS = density of states; EC/EV = conduction/valence band edges referenced to vacuum (EC = LUMO, EV = HOMO); rms = root-mean-square (AC amplitude reported as RMS) ); SRH = Shockley–Read–Hall traps, with trap density $N_t$ held constant, while electron and hole capture coefficients are varied to realize the bulk pseudo-lifetimes within the ranges given in Table 1. All optical constants were accessed via Fluxim AG SETFOS 5.5 material library; primary sources are cited in the reference list [O1]–[O5].

| **Layer / type** | **Parameter** | **Symbol** | **Value** | **Unit / Notes** |
|---|---|---|---|---|
| **Optical boundaries** | | — | **Air** | — |
| | Optical data | *n,k* | $n = 1.0$; $k = 0$ | incoherent |
| | Layer | — | Glass | — |
| | Thickness | t | 1 | mm |
| | Optical data | *n,k* | $n = 1.5$; $k = 0$ | incoherent |
| **Front electrode** | | — | **FTO** | — |
| | Thickness | t | 350 | nm |

| | Work function | Φ | 4.249 | eV; injection Ohmic |
|---|---|---|---|---|
| | Optical data | *n,k* | *Wenger et al., 2009 (via SETFOS 5.5)*[2] | imported *nk* file |
| **ETL** | | — | **$TiO_2$** | — |
| | Thickness | t | 50 | nm |
| | Energy levels (vacuum ref.) | $E_C$ / $E_V$ | 4.00 / 5.60 | eV |
| | Relative permittivity | $\varepsilon_r$ | 20.0 | — |
| | Effective DOS | $N_C$ / $N_V$ | $1.0\times10^{21}$ / $1.0\times10^{21}$ | $cm^{-3}$ |
| | Doping (donor / acceptor) | $N_D$ / $N_A$ | $1.1\times10^{20}$ / 0 | $cm^{-3}$ (n-type) |
| | Mobilities (constant) | $\mu_n$ / $\mu_p$ | 0.20 / $1.0\times10^{-10}$ | $cm^2.V^{-1}.s^{-1}$ |
| | Optical data | *n,k* | J.A. Woollam WVASE® dataset ($TiO_2$) | imported *nk* file |
| **Absorber** | | — | MAPI ($CH_3NH_3PbI_3$) | — |
| | Thickness | t | 370 | nm |
| | Energy levels (vacuum ref.) | $E_C$ / $E_V$ | 3.77 / 5.40 | eV |
| | Relative permittivity | $\varepsilon_r$ | 24.1 | — |
| | Effective DOS | $N_C$ / $N_V$ | $5.8\times10^{18}$/ $8.1\times10^{18}$ | $cm^{-3}$ |
| | Mobilities (constant) | $\mu_n$ / $\mu_p$ | 19 / 19 | $cm^2.V^{-1}.s^{-1}$ |
| | Optical generation | — | Enabled in absorber | — |
| | Bimolecular recombination | B | $9.4\times10^{-10}$ | $cm^3.s^{-1}$ |
| | SRH traps | Definition | Dirac, acceptor-like at $E_C$ - 0.8 | eV |
| | Trap density | $N_t$ | $1.0\times10^{16}$ | $cm^{-3}$ |
| | Optical data | n,k | *Löper et al., 2015 (via SETFOS 5.5)*[3] | imported nk file |
| **HTL** | | — | **Spiro-OMeTAD** | — |
| | Thickness | t | 220 | nm |
| | Energy levels (vacuum ref.) | $E_C$ / $E_V$ | 3.70 / 5.22 | eV |
| | Relative permittivity | $\varepsilon_r$ | 3.0 | — |
| | Effective DOS | $N_C$ / $N_V$ | $1.0\times10^{21}$ / $1.0\times10^{21}$ | $cm^{-3}$ |
| | Doping (donor / acceptor) | $N_D$ / $N_A$ | 0 / $1.1\times10^{20}$ | $cm^{-3}$ (p-type) |
| | Mobilities (constant) | $\mu_n$ / $\mu_p$ | $1.0\times10^{-10}$ / $1.0\times10^{-3}$ | $cm^2.V^{-1}.s^{-1}$ |
| | Optical data | n,k | *Filipič et al., 2015 (via SETFOS 5.5)*[4] | imported nk file<br>incoherent |
| **Back electrode** | | — | **Au** | — |
| | Thickness | t | 60 | nm |
| | Work function | Φ | 4.951 | eV; injection Ohmic |
| | Optical data | n,k | *Palik, 1985 (via SETFOS 5.5)*[5] | imported nk file |
| **Environment & illumination** | Temperature | T | 298 | K |
| | Spectrum | — | 1-sun | AM1.5G |
| | Optical sweep | — | 380–1000, step 1 | nm |
| **Bias conditions** | Open circuit (quasi) | DC bias | 1.0 | V |
| | Short circuit | DC bias | 0 | V |
| **AC (impedance) settings** | Amplitude (rms) | ΔV | 10 | mV |
| | Frequency range | $f_{min}$ – $f_{max}$ | 0.1 – $1\times10^6$ | Hz (log sweep) |
| | Sampling | — | 20 | points/decade |

## S2. Experimental procedures.

**Device fabrication:** The fluorine-doped tin oxide (FTO) substrates were initially brushed with a 2:98 vol% solution of Hellmanex in water and rinsed with deionized water. Subsequently, the substrates underwent sequential ultrasonic cleaning, spending 15 minutes each in Hellmanex solution, deionized water, acetone, and isopropanol. After cleaning, the substrates were dried under a stream of nitrogen gas and treated with UV/ozone for 15 minutes using an Ossila Ozone

Cleaner. A compact $TiO_2$ (c-$TiO_2$) layer with a thickness of 30–40 nm was deposited by spray pyrolysis. The precursor solution was prepared by mixing 1 mL of titanium diisopropoxide bis(acetylacetonate) (75% in 2-propanol) with 14 mL of absolute ethanol, which was immediately sprayed onto pre-heated substrates at 450 °C for 30 minutes using oxygen as the carrier gas. To facilitate subsequent gold electrode deposition, a small area of the substrate was masked with a glass slide during the spray process. After deposition, the substrates were allowed to cool to room temperature and subjected to an additional UV/ozone treatment for 15 minutes.

A mesoporous $TiO_2$ (m-$TiO_2$) solution was then prepared by mixing 150 mg of commercial $TiO_2$ paste (18NRT) with 1 mL of absolute ethanol and stirring overnight. This mixture was spin-coated onto the c-$TiO_2$ layer at 4000 rpm for 10 seconds. The resulting films were heated on a hot plate at 120 °C during deposition, followed by ramping to 450 °C and maintaining this temperature for 30 minutes before cooling. The FTO/c-$TiO_2$/m-$TiO_2$ substrates then underwent another UV/ozone treatment for 15 minutes before being transferred to a nitrogen glovebox ($O_2$ and $H_2O$ levels < 0.5 ppm; temperature 25–28 °C) for perovskite deposition.

The perovskite (PSK) layer was deposited inside the glovebox using the antisolvent technique. The precursor solution was prepared by dissolving methylammonium iodide (MAI) and lead(II) iodide ($PbI_2$) in a 1.3:1.3 molar ratio in a solvent mixture of 90% dimethylformamide (DMF) and 10% dimethylsulfoxide (DMSO) by volume. The solution was heated to 60 °C until fully dissolved, then filtered through a 0.45 μm PTFE syringe filter. A 100 μL aliquot of this precursor solution was spin-coated onto the $TiO_2$ films at 3000 rpm for 30 seconds. Eight seconds after the start of spinning, 200 μL of ethyl acetate (EA) was dropped onto the center of the spinning substrate. Immediately after spin-coating, the films were placed on a hot plate at 65 °C for 1 minute, followed by annealing at 100 °C for 40 minutes.

The hole transport layer was deposited by spin-coating a 0.07 M solution of Spiro-OMeTAD in chlorobenzene, which was doped with LiTFSI (1.8 M in acetonitrile), FK209 Co(III) (0.25 M in acetonitrile), and 4-tert-butylpyridine, in molar ratios of 0.5, 0.03, and 3.3, respectively. The solution was filtered before spin-coating in two consecutive steps: first at 2000 rpm for 5 seconds, followed by 4000 rpm for 30 seconds, both at room temperature. Finally, gold electrodes were thermally evaporated through a shadow mask under high vacuum ($\sim 10^{-6}$ mbar), completing the device architecture.

**Device characterization:** Scanning electron microscopy (SEM) micrographs were obtained using a Hitachi S4800 microscope operated at 2 kV to determine the thickness and uniformity of the layers in cross-section (**Figure S1**).

Current-voltage (I-V) curves of the fabricated devices were recorded under a solar simulator (ABET-Sun2000) equipped with an AM 1.5 G filter at 100 $mW/cm^2$ under 1 sun illumination. The measurements were performed in reverse scan with a scan rate of 100 mV/s from 1.2 to -0.1 V. The samples were measured with a black mask having an active area of 0.14 $cm^2$ to calculate photovoltaic (PV) parameters such as open circuit voltage ($V_{OC}$), short-circuit current density ($J_{SC}$), fill factor (FF), and power conversion efficiency (PCE).

Electrochemical impedance spectroscopy (EIS) measurements were carried out using a Metrohm Autolab Multi Autolab PGSTAT204 system equipped with a FRA32M module. A WAVELABS SINUS-300 solar simulator, calibrated to 1 sun (AM1.5G), was used as the light source. The measurements were performed at applied voltages of 1 V (quasi–open circuit) and 0 V (short circuit). The AC perturbation amplitude was set to 10 mV (RMS). Impedance spectra were recorded over a frequency range from 1 MHz to 0.1 Hz, using 140 frequencies with a logarithmic step. All measurements were performed at 24.5 °C under an inert nitrogen atmosphere using a 0.14 $cm^2$ shadow mask. Data acquisition and analysis were performed using NOVA 2.1 software.

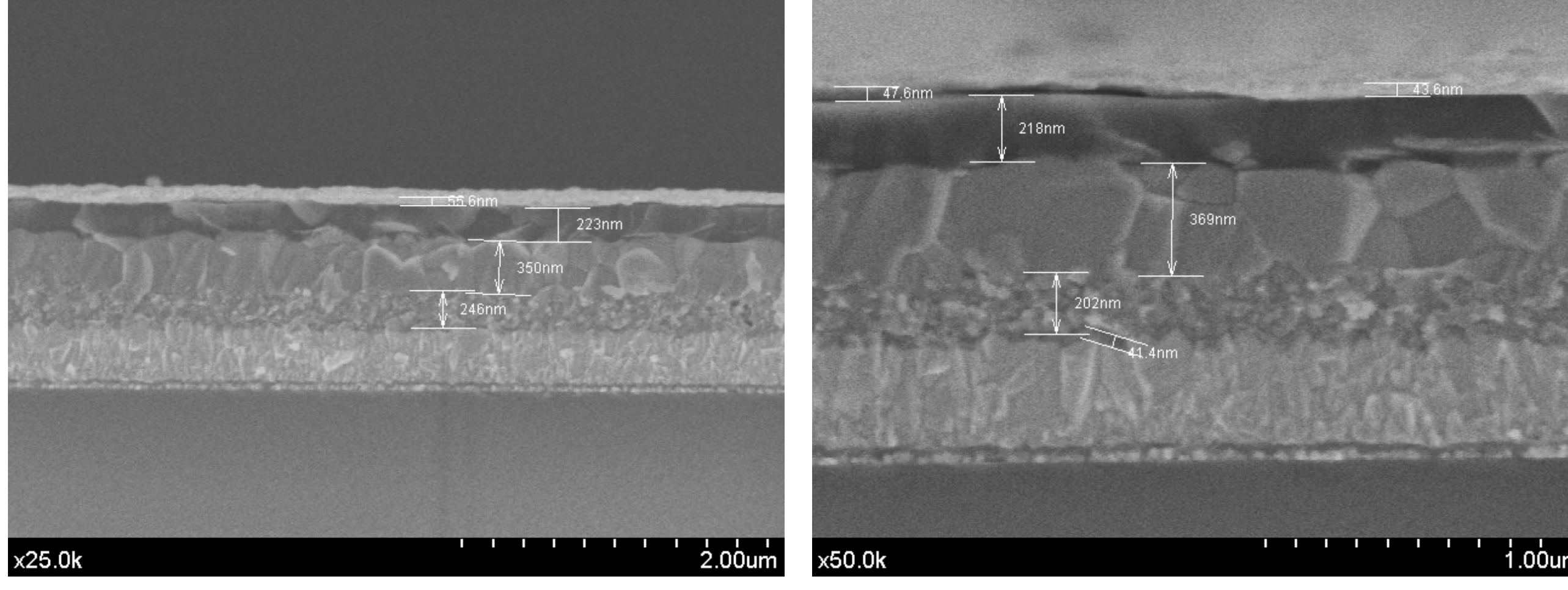

**Figure S1.** Cross-section SEM of FTO/c-$TiO_2$/m-$TiO_2$/MAPI/Spiro/Au devices at ×25k (left) and ×50k (right). Representative thicknesses: Au ≈ 44–56 nm, Spiro ≈ 218–223 MAPI ≈ 350–369 nm, m-$TiO_2$ ≈ 245 nm, c-$TiO_2$ ≈ 42 nm, and FTO ≈ 350 nm. For simplicity, the simulations use a planar c-$TiO_2$/MAPI/Spiro stack for the DD/EIS simulations (Figure 2). Although the experimental devices include a m-$TiO_2$ scaffold infiltrated with perovskite, electronic transport through this scaffold is negligible. Its influence is instead captured through $TiO_2$/perovskite interfacial parameters, which provides a more appropriate and efficient representation in a one-dimensional drift–diffusion model

## S3. ML-based analysis.

**Table S2.** Codes used to refer to features of the EIS spectra.

**Equivalent Circuit Fitting (ECF):**

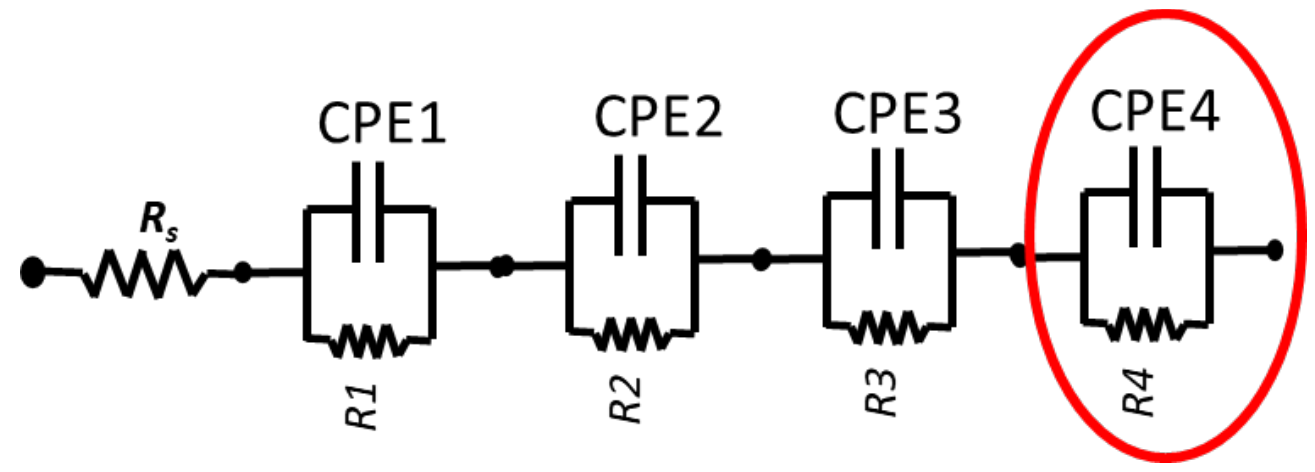

| Feature code | Meaning | Units |
|---|---|---|
| Rs* | Series resistance | Ω $cm^2$ |
| R1 | High frequency resistance | Ω $cm^2$ |
| R2 | Medium frequency resistance | Ω $cm^2$ |
| R3_clean | Low frequency resistance after processing (excluding infinite values) | Ω $cm^2$ |
| R3_switch | Binary feature: 0 for infinity low frequency resistance, 1 otherwise | No units |
| G3 | Low frequency conductance | $\Omega^{-1}$ $cm^{-2}$ |
| CP1-T, CP2-T, CP3-T | High, medium and low frequency capacitance, respectively | Ω / $cm^2$ |
| CP1-P, CP2-P, CP3-P | High, medium and low frequency capacitor ideality factor, respectively | No units |
| R4, CP4-T, CP4-T | Additional RC element (OC only), when the low-frequency signal exhibited a change of sign in the imaginary component | |

*Rs was included in the fits but not included in the ML training.

**Direct signal featurization (DSF):**

| Feature code | Meaning | Units |
|---|---|---|
| f1, f2 | Frequency positions of the high and low frequency peak, respectively | Hz |
| h1, h2 | Height of the high and low frequency peak, respectively | Ω $cm^2$ |
| fwhm1, fwhm2 | FWHM of the high and low frequency peak, respectively | Ω $cm^2$ |
| r0 | Low frequency limit (0.1 Hz) of the real impedance | Ω $cm^2$ |
| i0 | Low frequency limit (0.1 Hz) of the real impedance | Ω $cm^2$ |

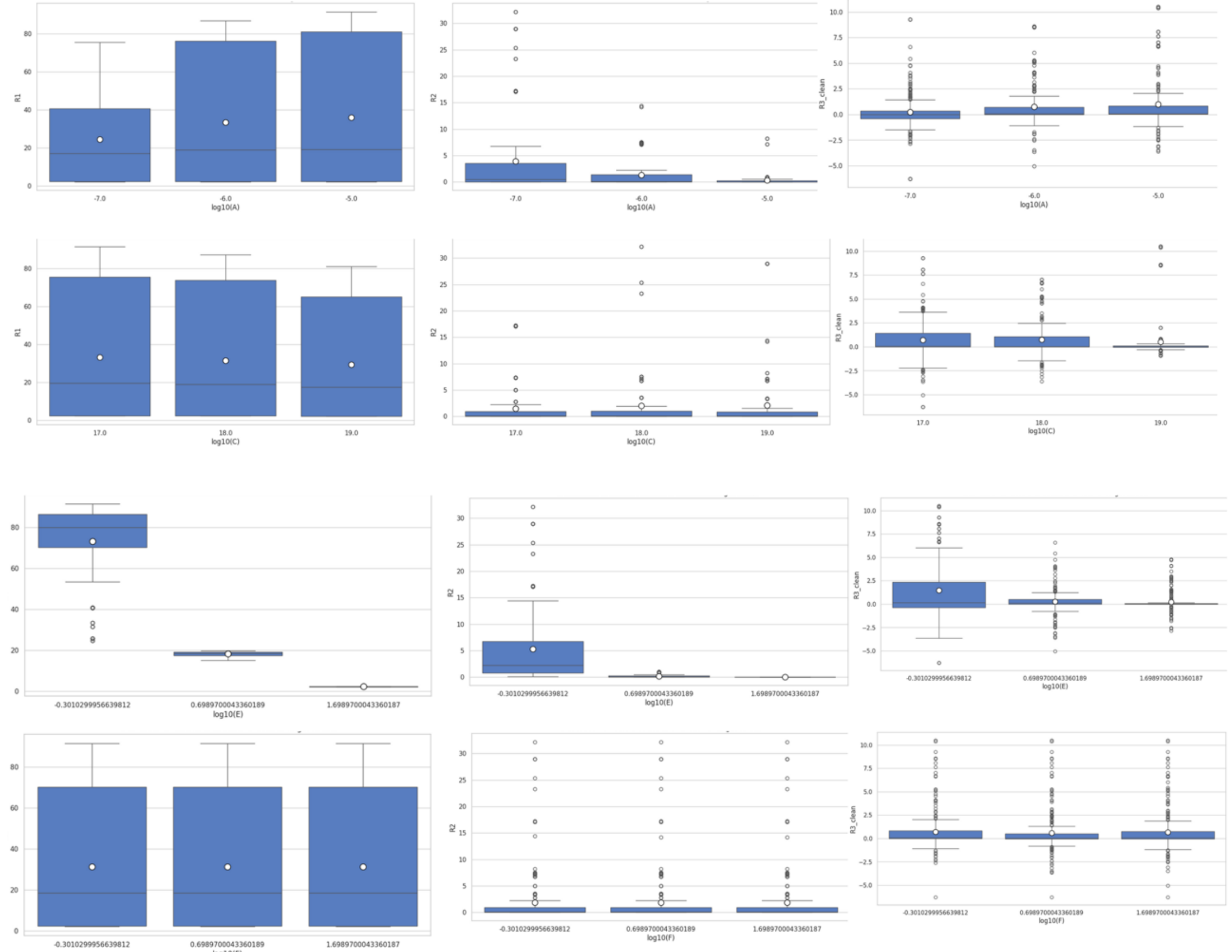

**Figure S2.** Histograms of ECF features for the DD simulations run for a set of $3^6 = 729$ spectra covering a regular distribution of targets under OC conditions. Top 6 figures show the evolution of R1, R2 and "R3-clean" with respect to A and C parameters. Bottom 6 figures show the evolution of the same features with respect to E and F' parameters. The modified F' parameter represents the surface recombination velocity of electrons (majority carrier) at the $TiO_2$ interface.

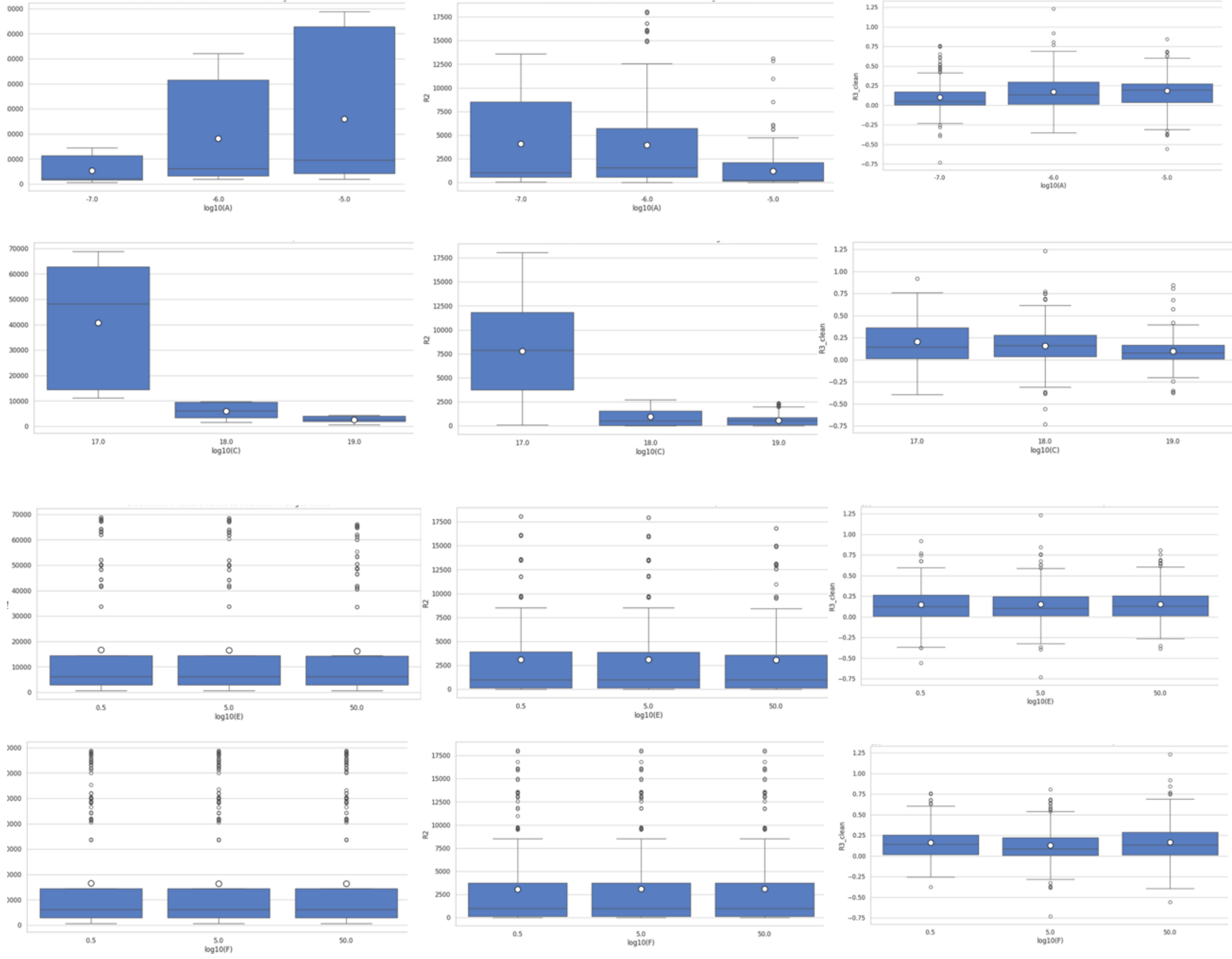

**Figure S3.** Same as Figure S3 for the SC case.

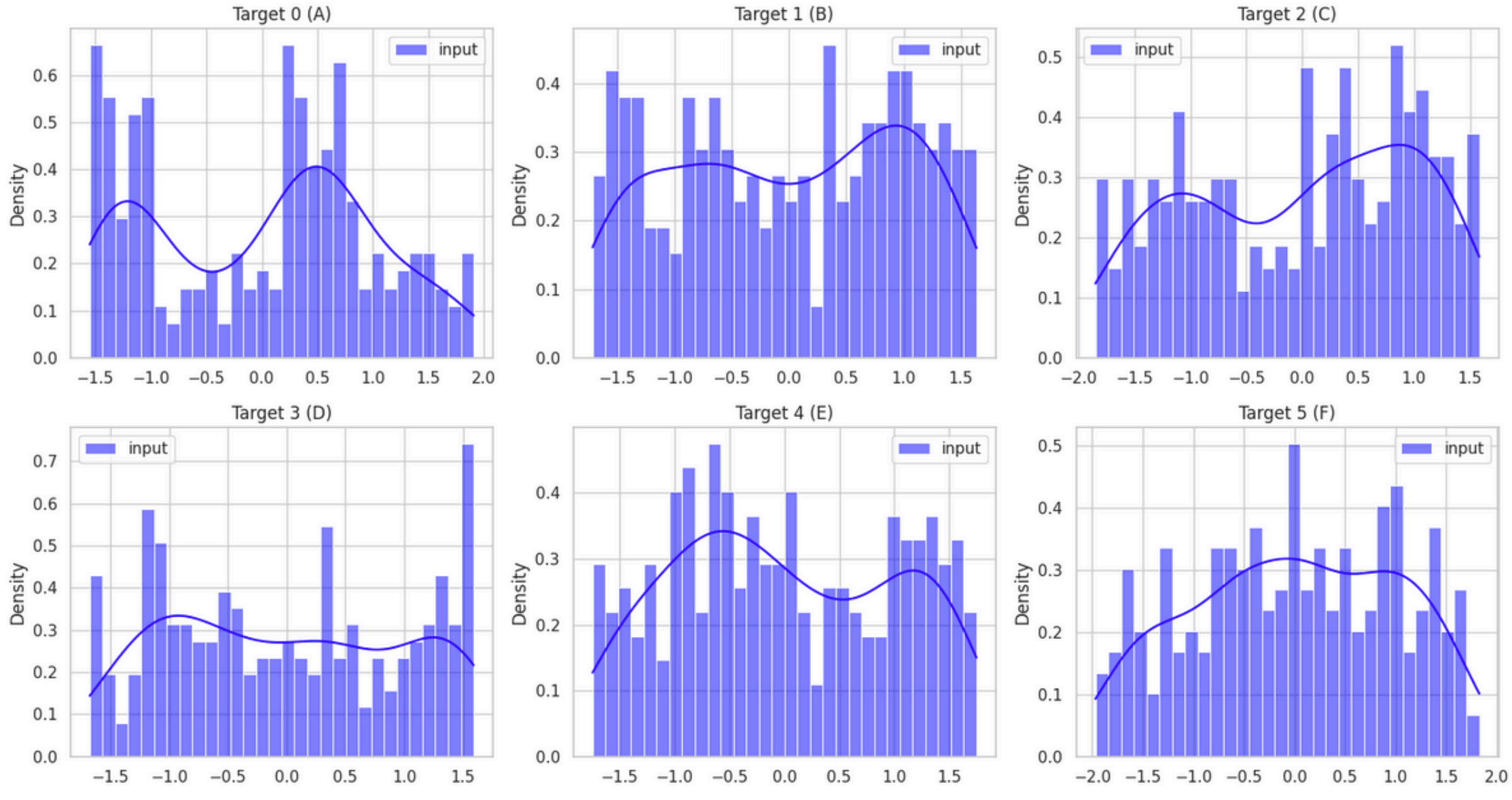

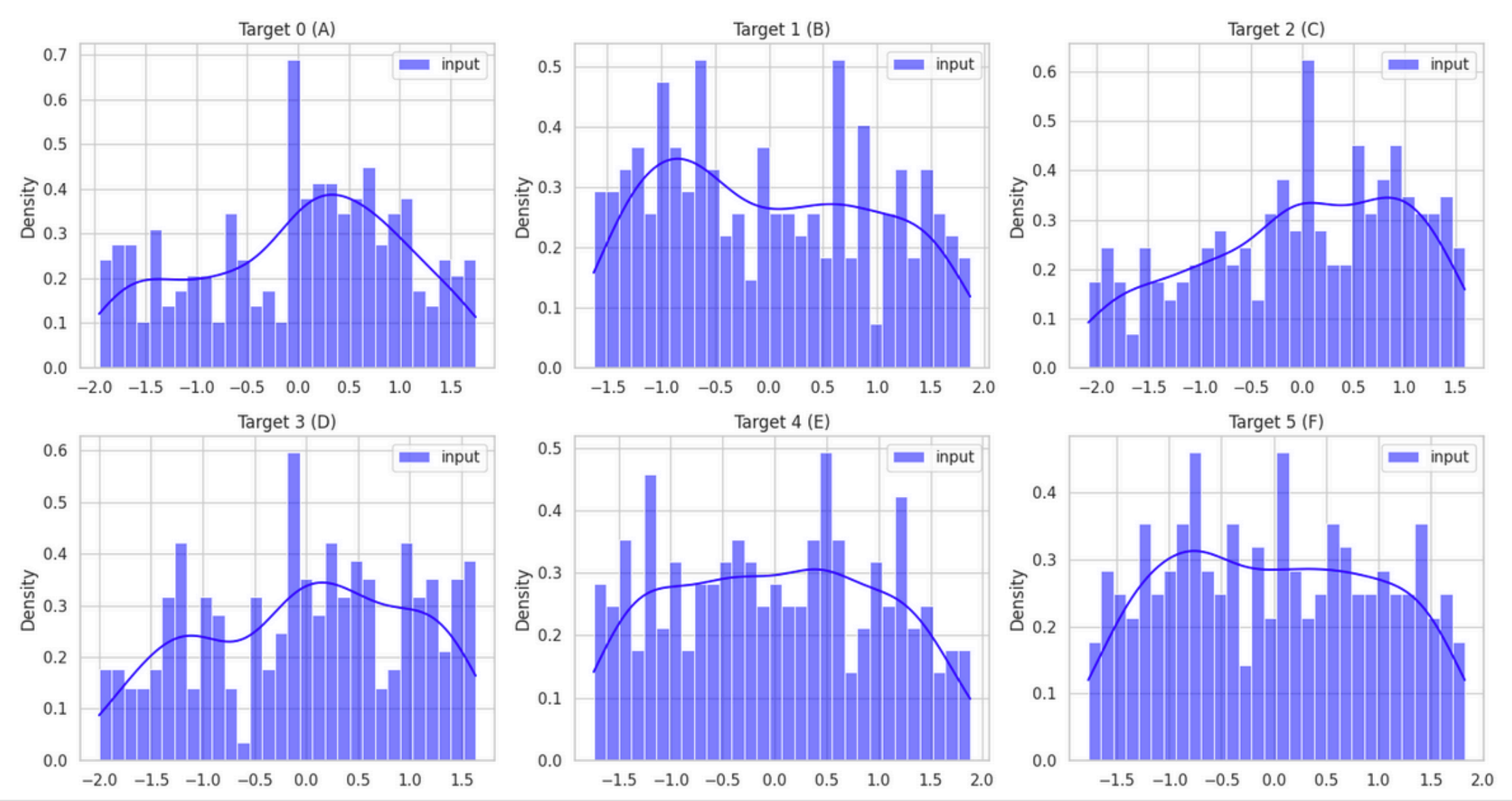

**Figure S4.** Distribution of targets in the random sweeps. OC (top), SC (bottom).

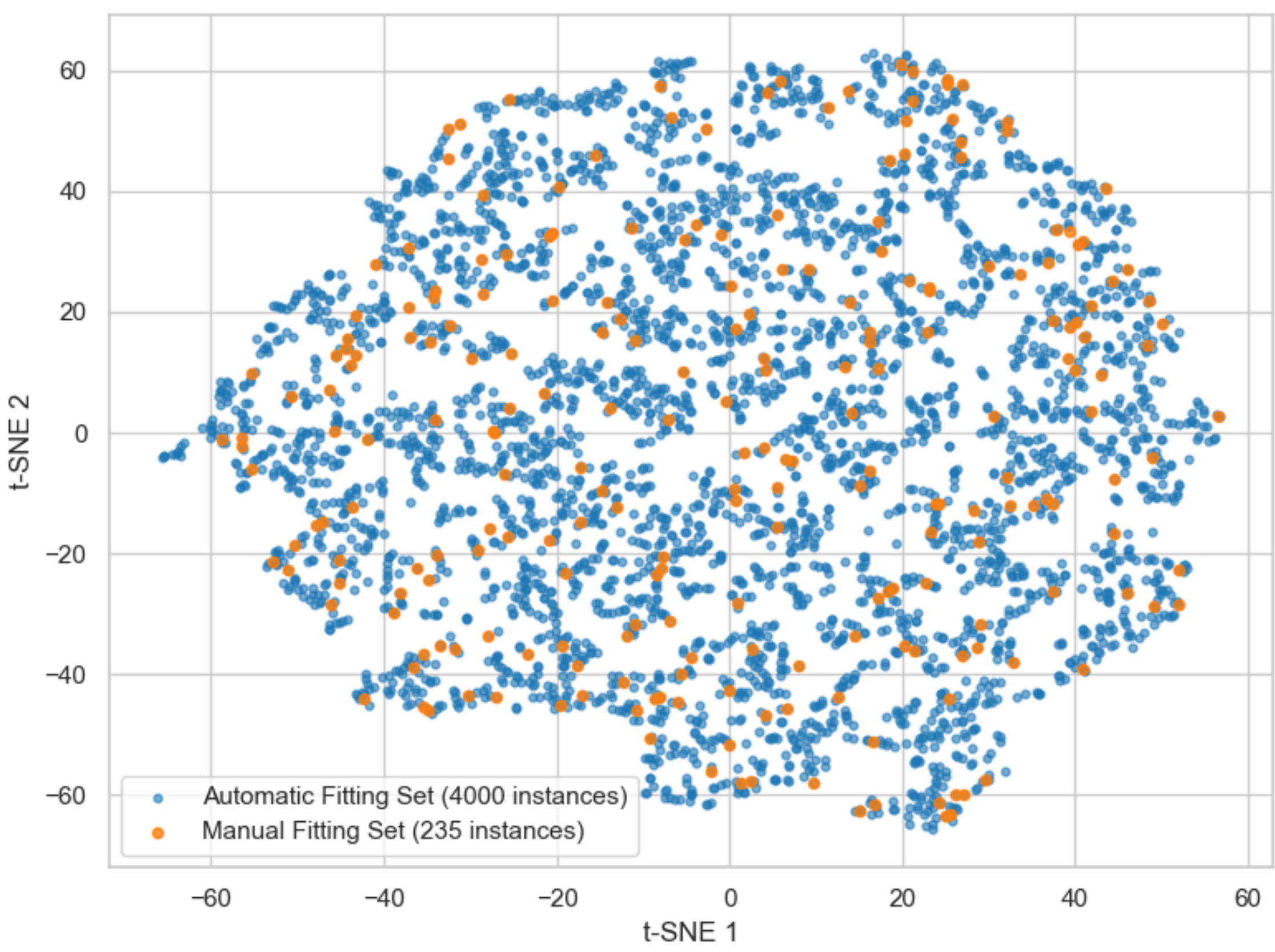

**Figure S5.** Distribution of instances used for training in both the original (manual) and the expanded (automatic) sets

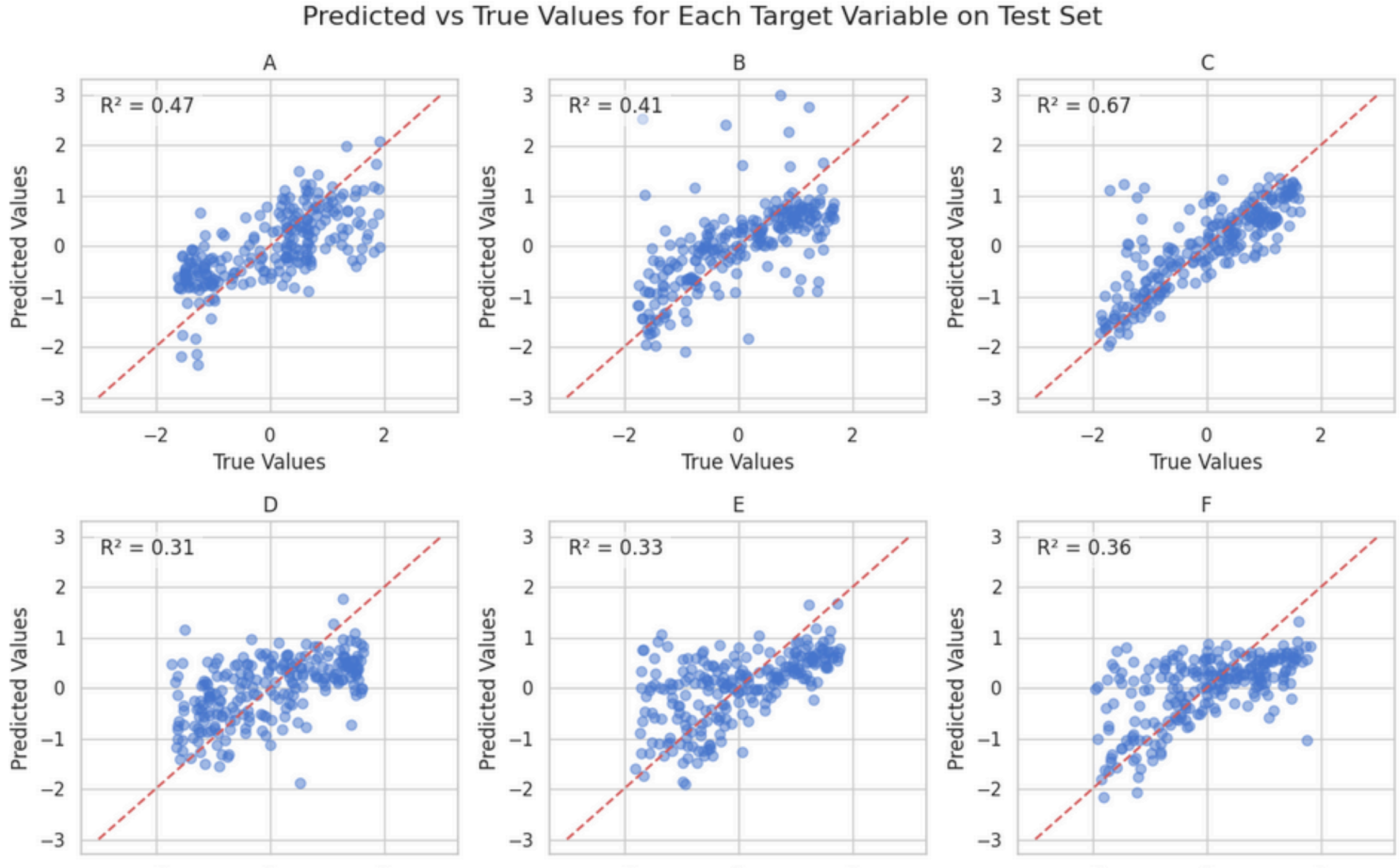

**Figure S6.** Predicted versus true values for each target variable for the MLP model using ECF featurization under OC conditions. Note that normalized or standardized values are shown in the plots.

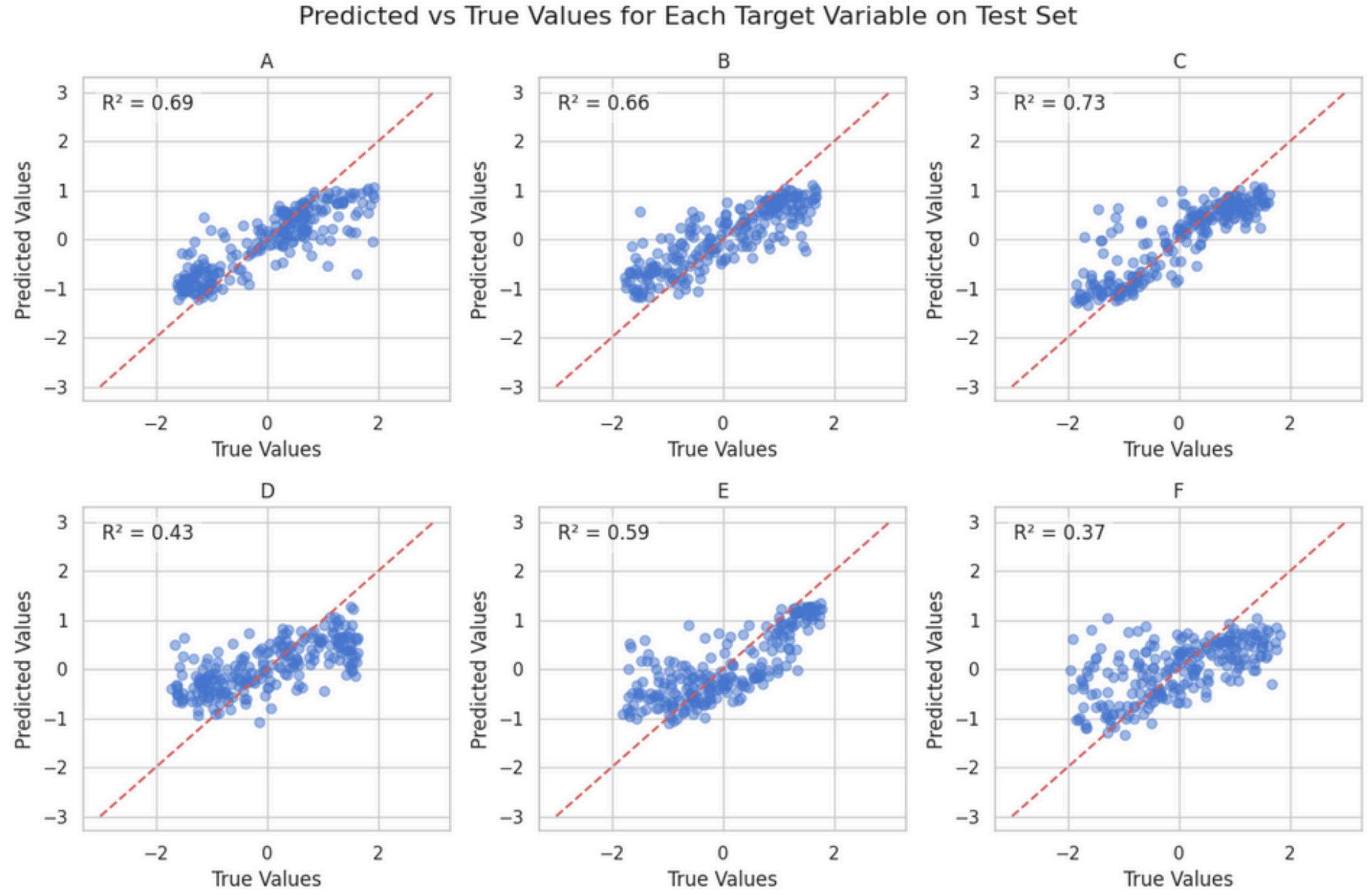

**Figure S7.** Same as Figure S6 for the RF model.

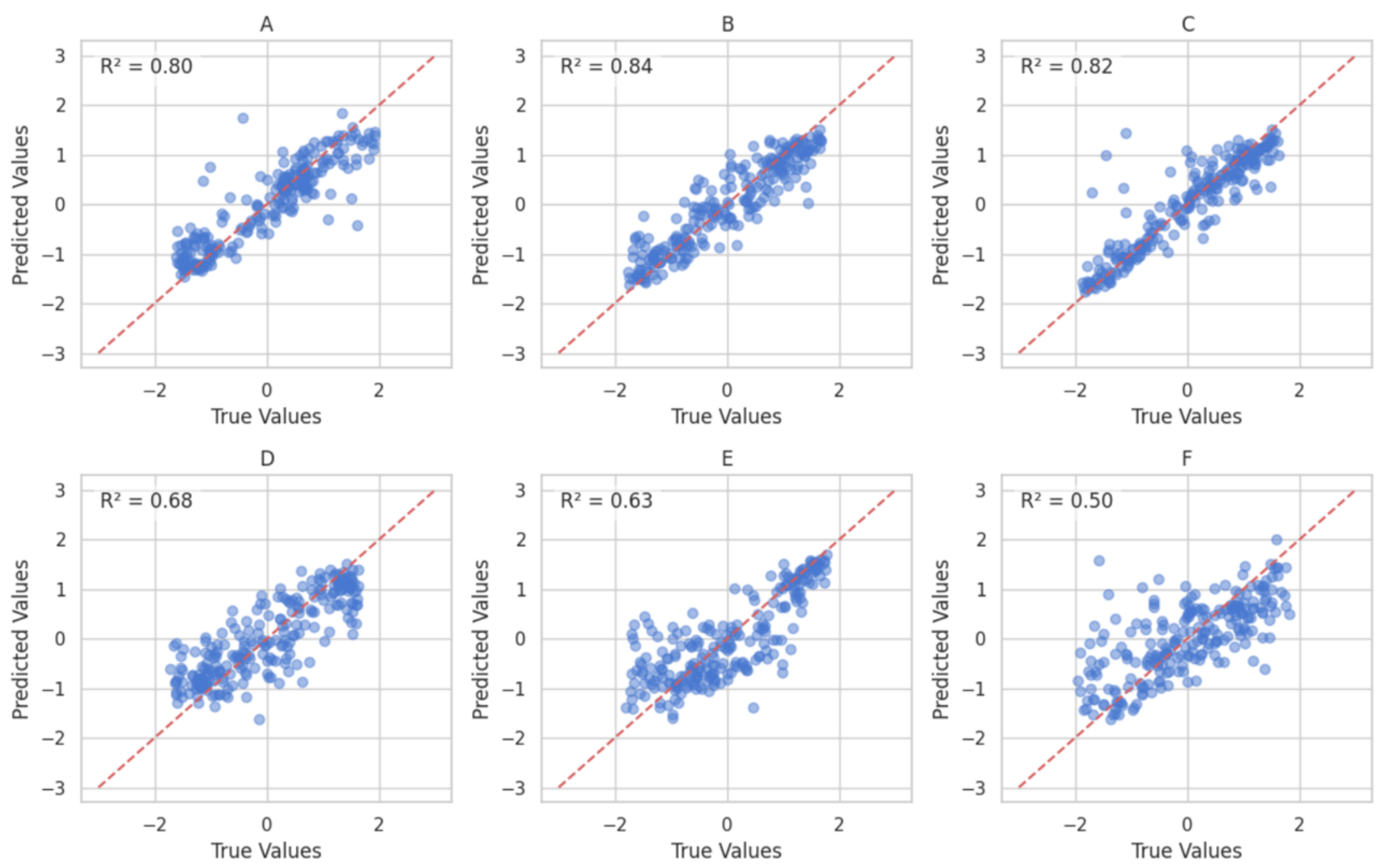

**Figure S8.** Same as Figure S6 for the GBR model.

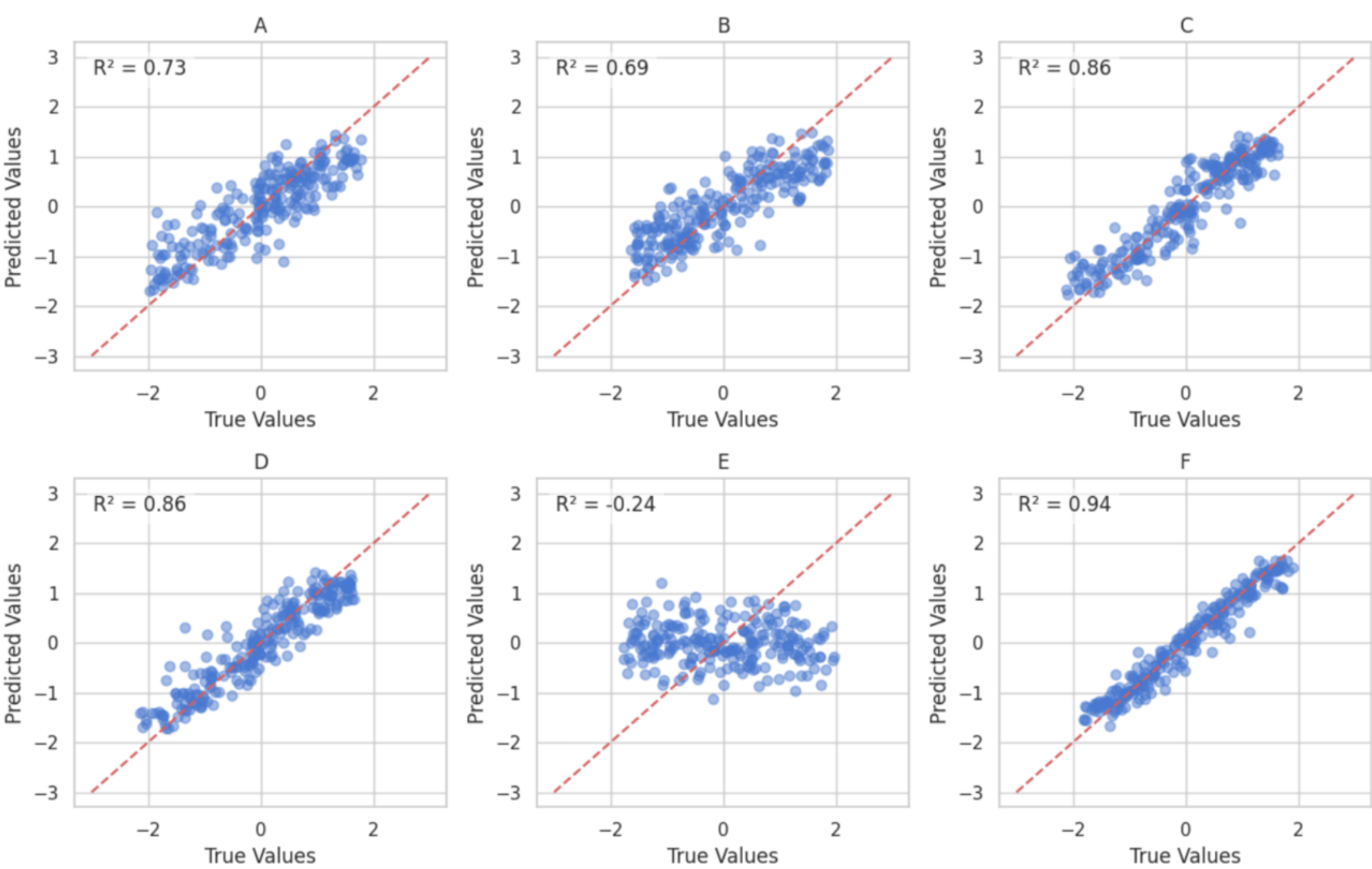

**Figure S9.** Predicted versus true values for each target variable for the GBR model using ECF featurization for the original 235-sized set under SC conditions.

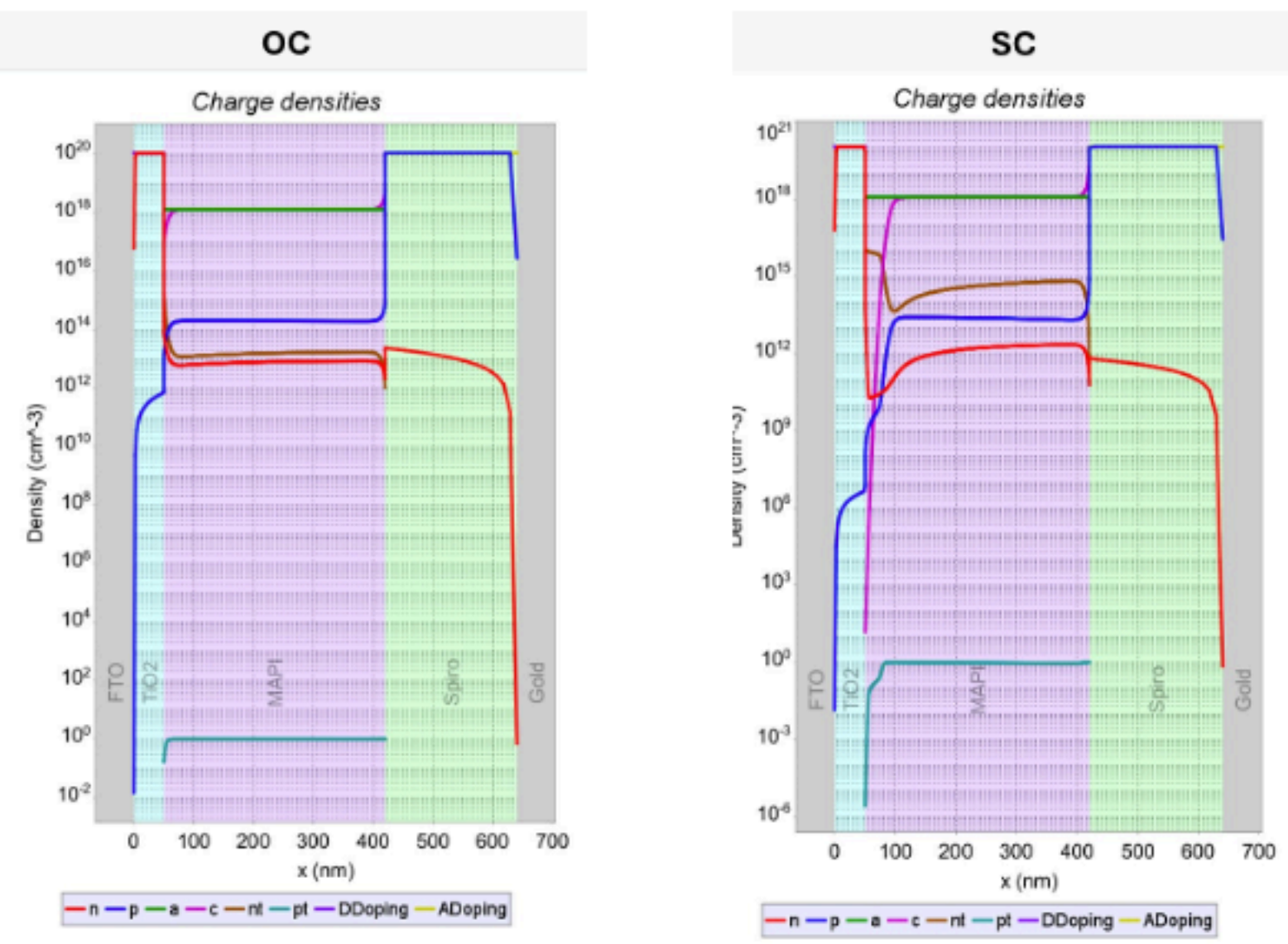

**Figure S10.** Carrier density profiles under OC and SC conditions as obtained by the DD numerical model (SETFOS)

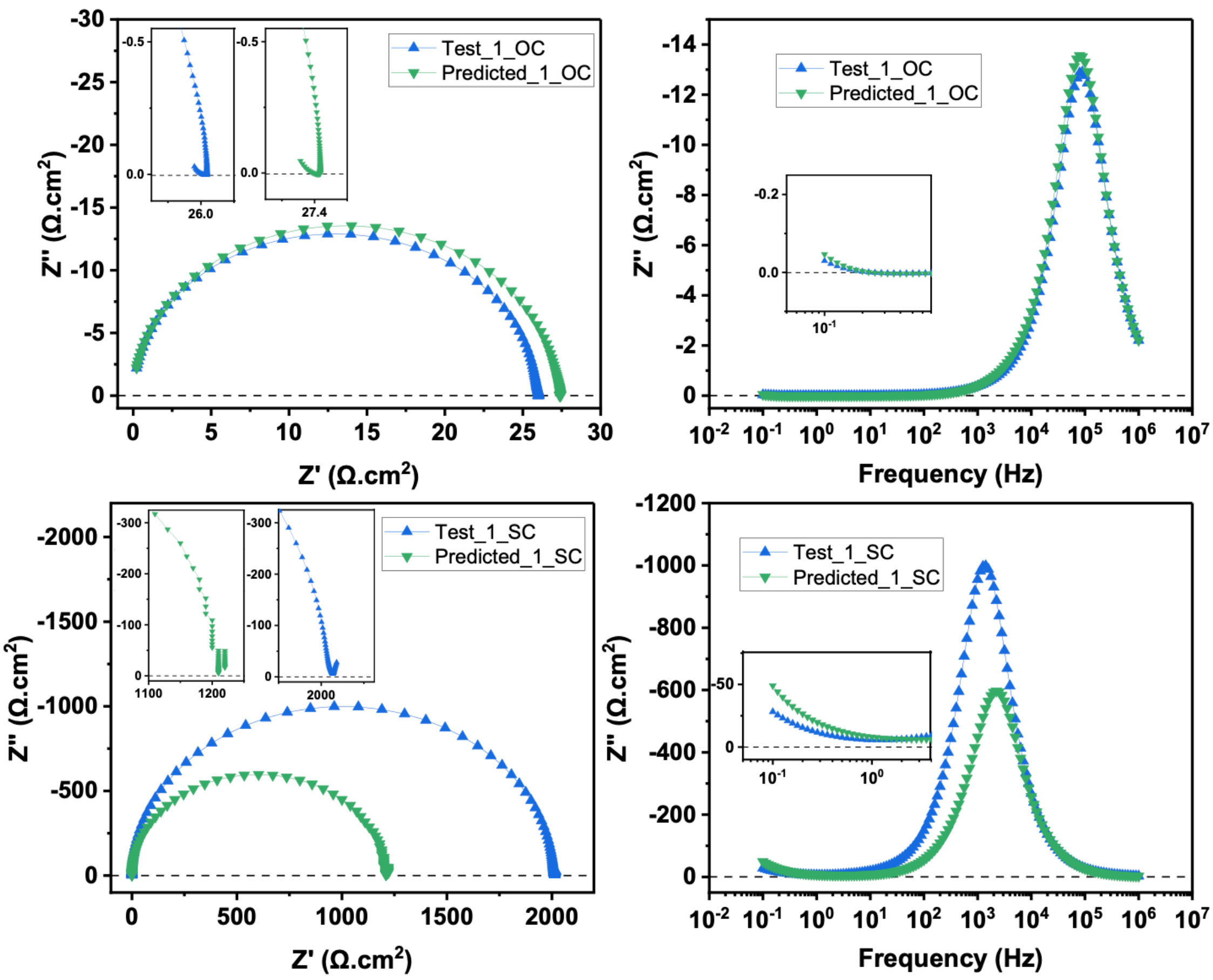

**Figure S11.** Predicted versus true spectra for the random instance not included neither in the training nor the test set. Results under OC (top) and SC (bottom) conditions are shown.

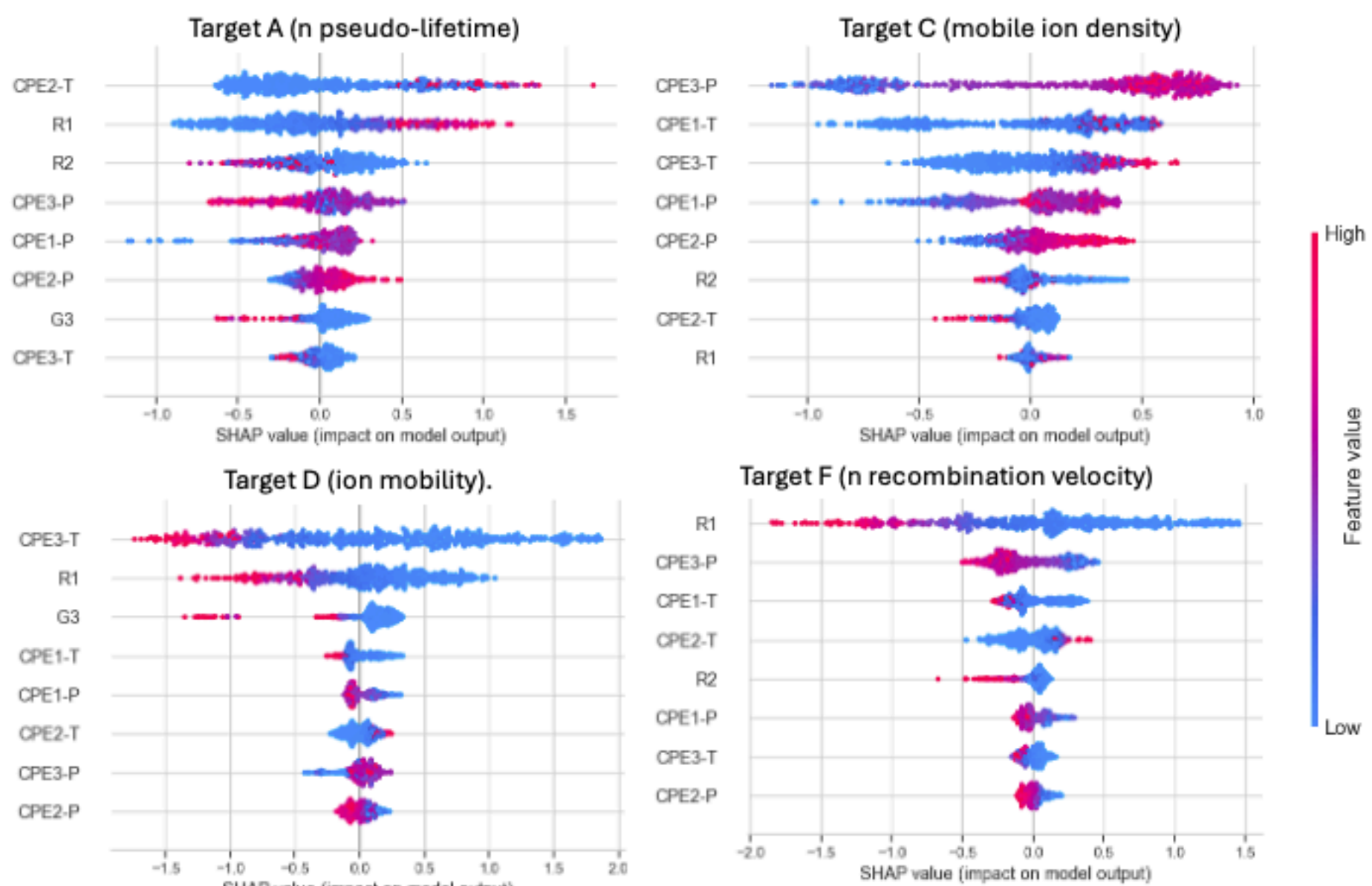

**Figure S12.** SHAP analysis for the GBR results under SC conditions with ECF featurization. SHAP analysis are interpreted in the following way: features are listed in order of importance. The SHAP value shows the impact of that feature on the prediction of each target. A positive value pushes the prediction higher, a negative value pushes the prediction lower. Red colour means high feature value for that instance. Blue colour means low feature value for that instance. Feature codes: R*x*'s = resistances; CPE*x*-T's = capacitances, CPE*x*-P's = capacitor ideality factors, where $x$ = 1,2,3,4 is counter running from high to low frequencies. G3= low frequency conductance.

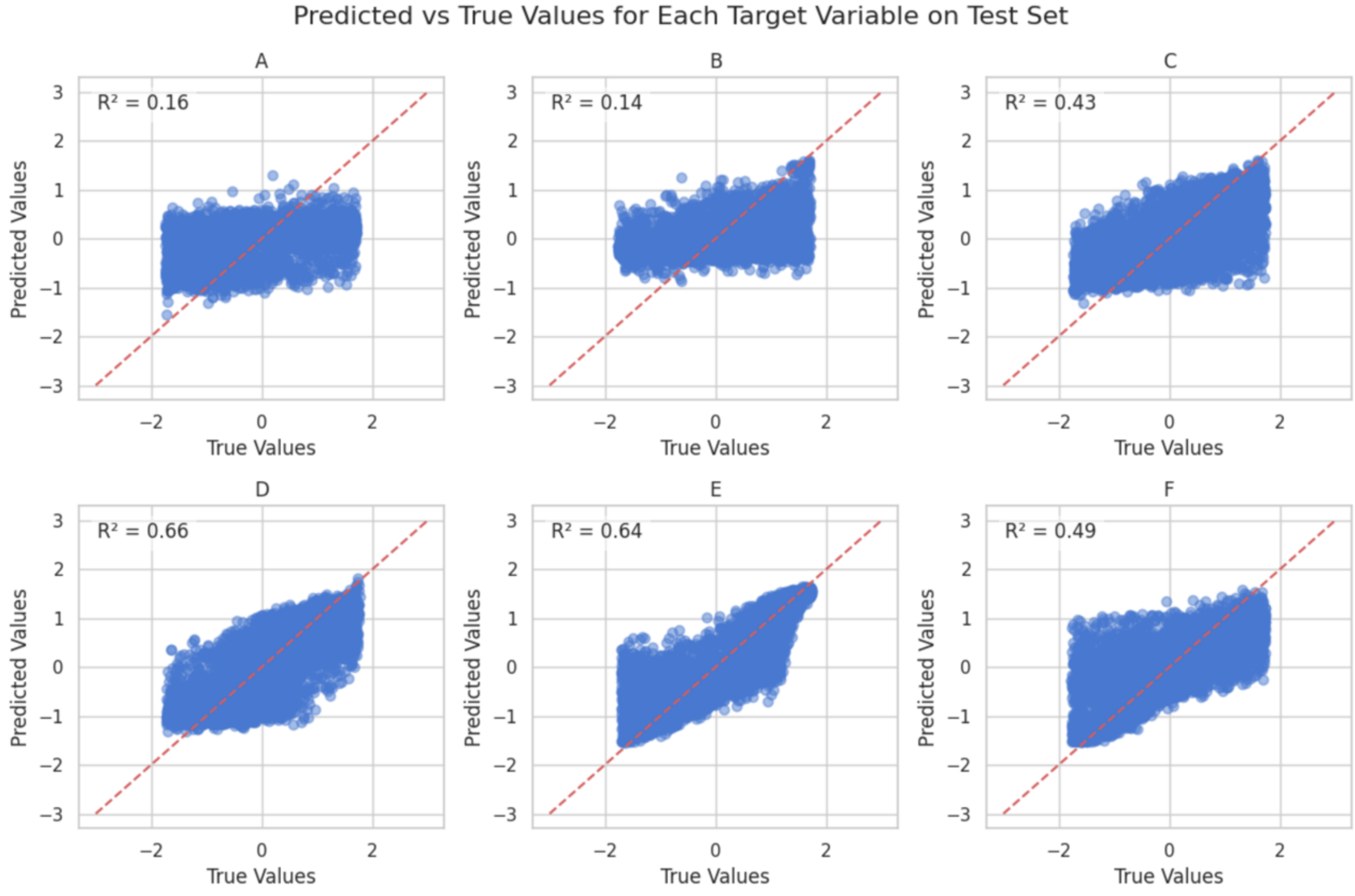

**Figure S13.** Predicted versus true values for each target variable for the GBR model using DSF featurization under OC conditions. Note that normalized or standardized values are shown in the plots

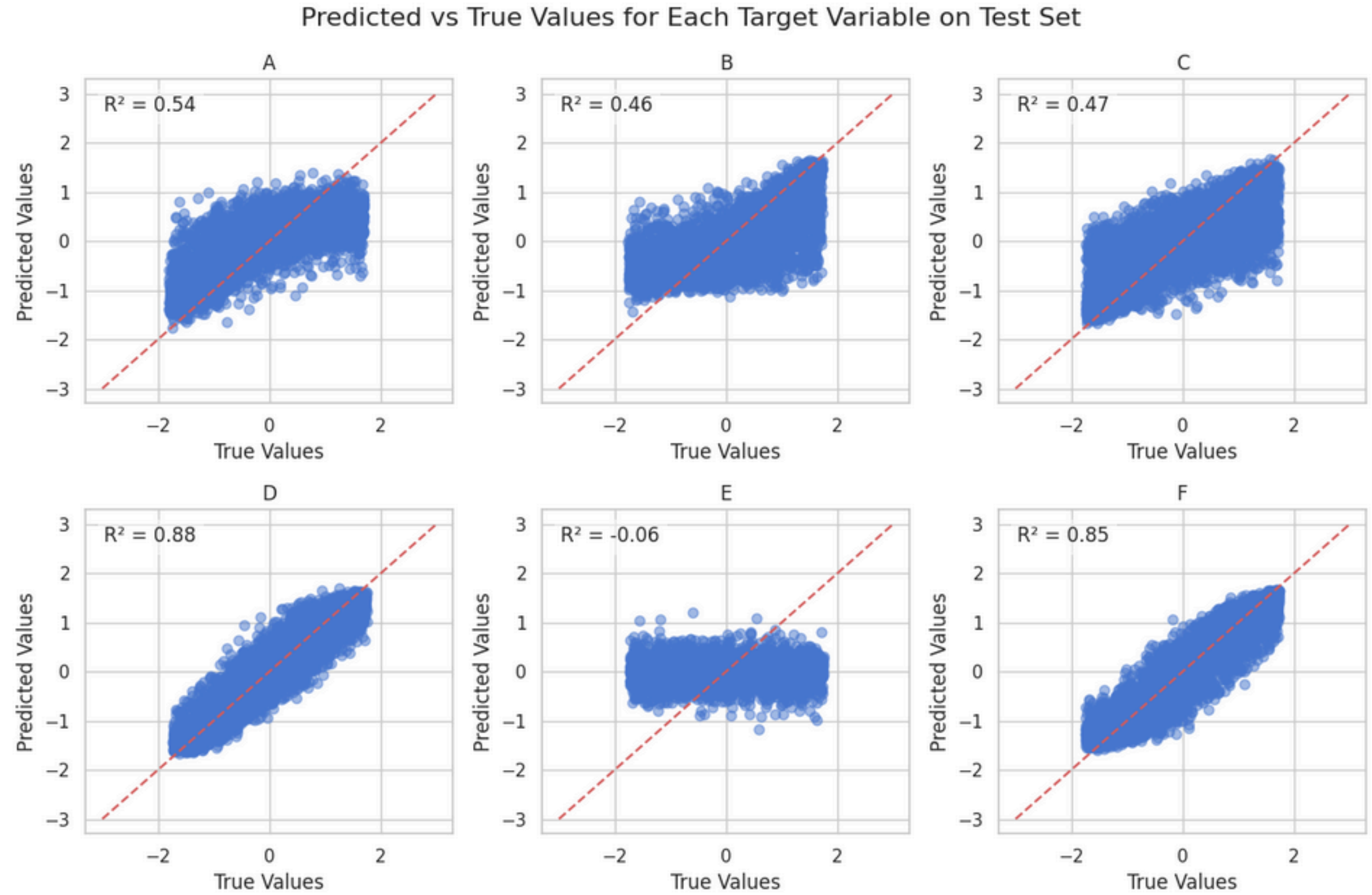

**Figure S14.** Same as S13 under SC conditions.

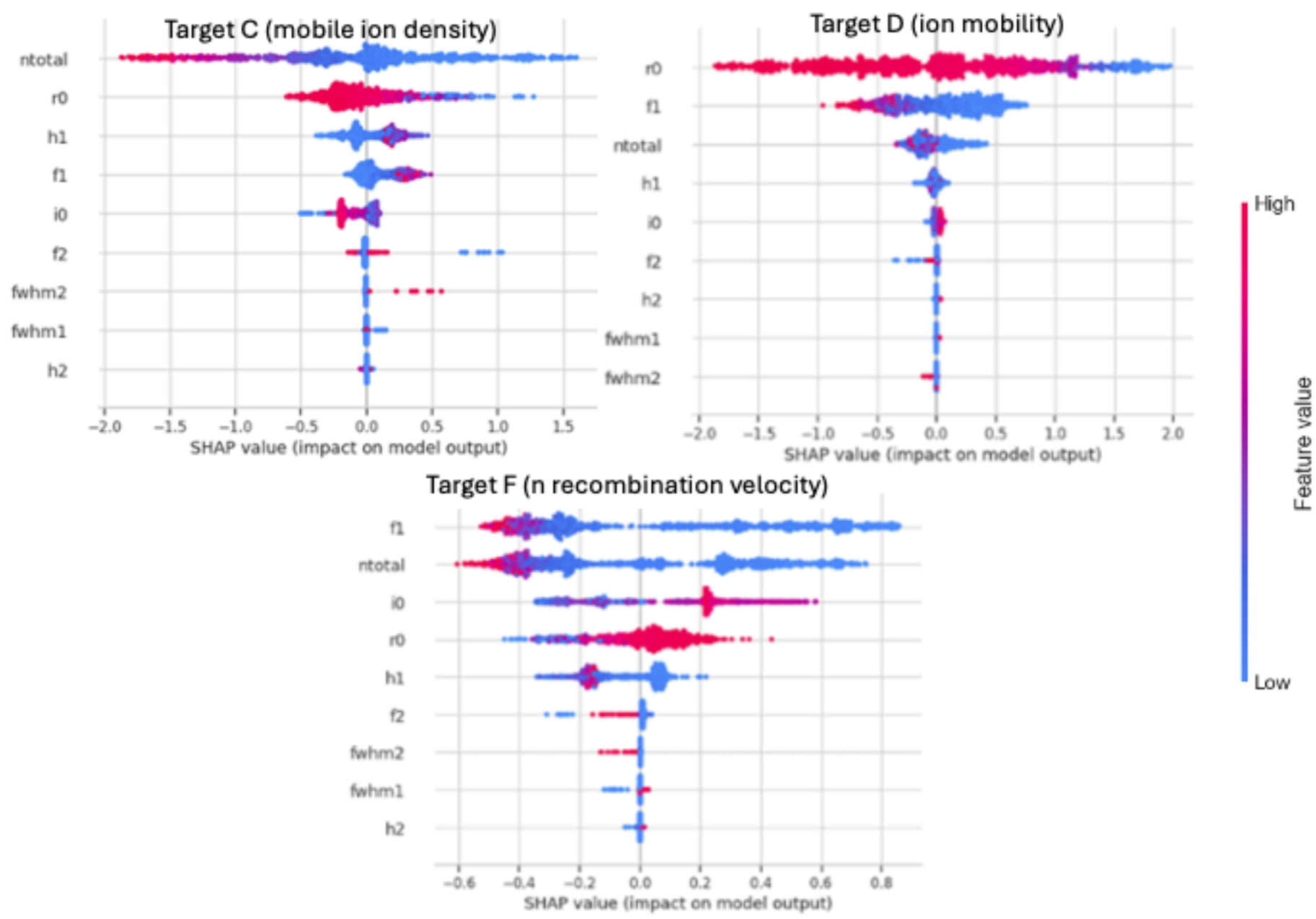

**Figure S15.** SHAP analysis for the GBR results under SC conditions with features extracted by DSF. Check Table S2 for the meaning of the feature codes. Feature codes: f*x*'s = peaks frequency positions; h*x*'s = peaks heights, fwhm*x*'s = peaks widths, where *x* = 1,2 is counter running from high to low frequencies. r0, i0, low frequency limits of the real and the imaginary impedance, respectively. Ntotal = number of peaks found.

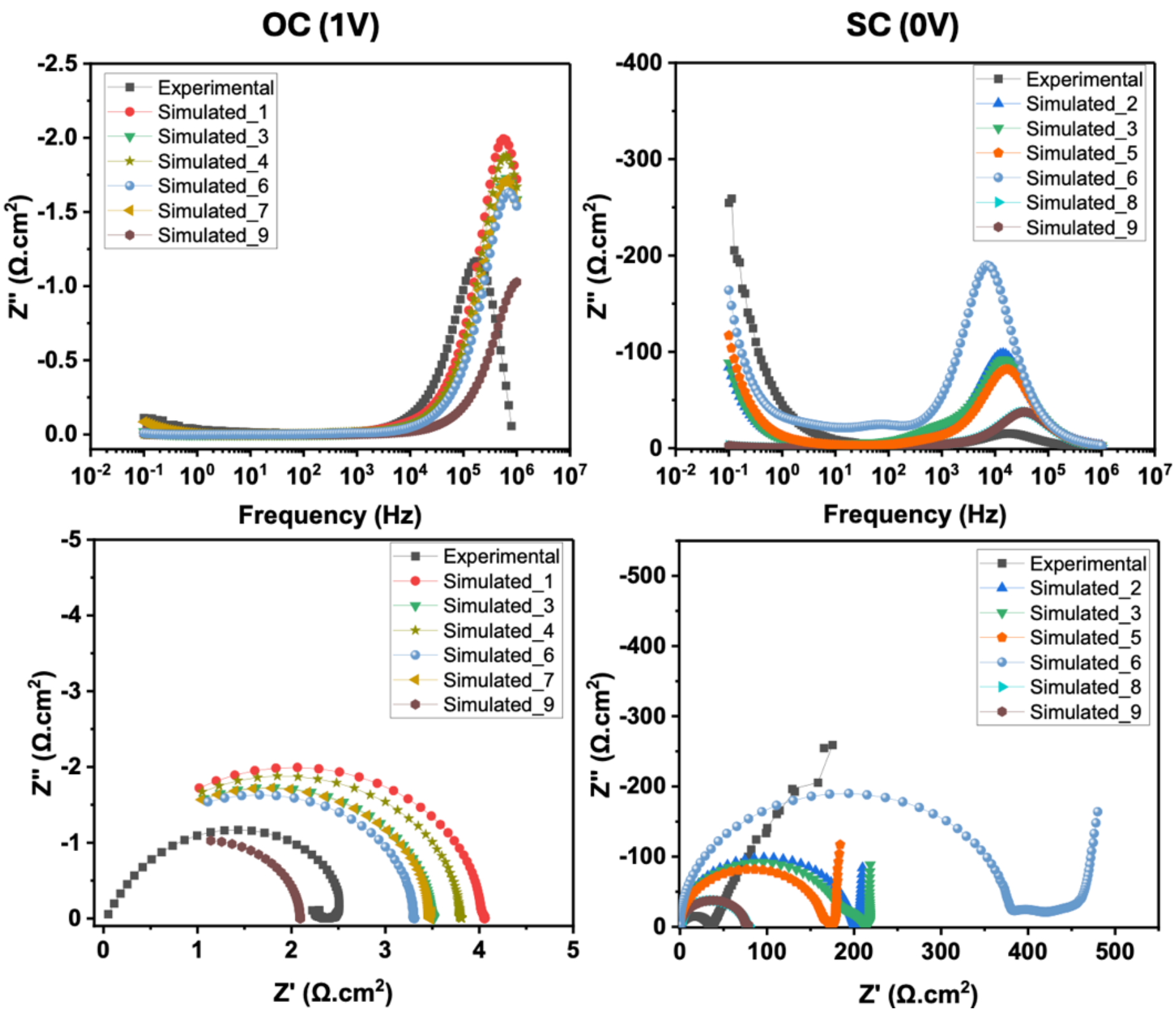

**Figure S16.** Comparison between experimental and simulated EIS spectra using the predicted values of Table 3.